\documentclass{aa}  
\bibpunct{(}{)}{;}{a}{}{,}
\usepackage{afterpage,natbib,lipsum}
\usepackage{graphicx}
\usepackage{multirow}
\usepackage{txfonts} 
\usepackage{hyperref}
\usepackage{graphicx}	% Including figure files
\usepackage{subfig}
\usepackage{amsmath}	% Advanced maths commands
\usepackage{amssymb}	% Extra maths symbols

\usepackage{color}
\usepackage{rotating}
\usepackage{url}
\usepackage{ifthen}

\usepackage{bm}
\usepackage{bbold}
\usepackage{aas_macros}
\usepackage{verbatim}
\usepackage{algorithm}% http://ctan.org/pkg/algorithms
\usepackage{algpseudocode}% http://ctan.org/pkg/algorithmicx
\usepackage{enumitem}

\DeclareMathAlphabet{\mathpzc}{OT1}{pzc}{m}{it}

\newcommand\Gtilde{\stackrel{\sim}{\smash{\mathcal{G}}\rule{0pt}{1.1ex}}}
\newcommand{\Mpch}{{$h^{-1}$~Mpc}}

\newcommand{\mvec}[1]{\bm{#1}}
\newcommand{\mmat}[1]{\mathbf{#1}}

\newcommand{\myvec}[1]{\mvec{#1}}
\newcommand{\mymat}[1]{\mmat{#1}}
\newcommand\numberthis{\addtocounter{equation}{1}\tag{\theequation}}
\makeatletter
\newcommand{\vast}{\bBigg@{3}}
\newcommand{\Vast}{\bBigg@{4}}
\makeatother
\usepackage[dvipsnames]{xcolor}
\hypersetup{
  draft = false,
  colorlinks   = true, %Colours links instead of boxes
  urlcolor     = RoyalBlue, %Colour for external hyperlinks
  linkcolor    = RoyalBlue, %Colour of internal links
  citecolor   = MidnightBlue %Colour of citations
}

\SetSymbolFont{symbols}{bold}{OMS}{cmsy}{b}{n}
\DeclareSymbolFont{bmisymbols}{OML}{cmm}{b}{it}

\begin{document} 

   \title{Cosmological inference from Bayesian forward modelling of deep galaxy redshift surveys}

   \author{Doogesh Kodi Ramanah\inst{1,2} \and Guilhem Lavaux\inst{1,2} \and Jens Jasche\inst{3} \and Benjamin D. Wandelt\inst{1,2,4} }

   \institute{Sorbonne Universit\'e, CNRS, UMR 7095, Institut d'Astrophysique de Paris, 98 bis bd Arago, 75014 Paris, France             
	\and
    Sorbonne Universit\'es, Institut  Lagrange  de  Paris  (ILP),  98  bis bd Arago, 75014 Paris, France
	\and
	The Oskar Klein Centre, Department of Physics, Stockholm University, AlbaNova University Centre,
SE 106 91 Stockholm, Sweden
    \and
    Center for Computational Astrophysics, Flatiron Institute, 162 5th Avenue, 10010, New York, NY, USA
   }

   \date{Received XXX; accepted XXX}

   \abstract{We present a large-scale Bayesian inference framework to constrain cosmological parameters using galaxy redshift surveys, via an application of the Alcock-Paczy\'nski (AP) test. Our physical model of the non-linearly evolved density field, as probed by galaxy surveys, employs Lagrangian perturbation theory (LPT) to connect Gaussian initial conditions to the final density field, followed by a coordinate transformation to obtain the redshift space representation for comparison with data. We implement a Hamiltonian Monte Carlo sampler to generate realizations of three-dimensional (3D) primordial and present-day matter fluctuations from a non-Gaussian LPT-Poissonian density posterior given a set of observations. This hierarchical approach encodes a novel AP test, extracting several orders of magnitude more information from the cosmic expansion compared to classical approaches, to infer cosmological parameters and jointly reconstruct the underlying 3D dark matter density field. The novelty of this AP test lies in constraining the comoving-redshift transformation to infer the appropriate cosmology which yields isotropic correlations of the galaxy density field, with the underlying assumption relying purely on the geometrical symmetries of the cosmological principle. Such an AP test does not rely explicitly on modelling the full statistics of the field. We verify in depth via simulations that this renders our test robust to model misspecification. This leads to another crucial advantage, namely that the cosmological parameters exhibit extremely weak dependence on the currently unresolved phenomenon of galaxy bias, thereby circumventing a potentially key limitation. This is consequently among the first methods to extract a large fraction of information from statistics other than that of direct density contrast correlations, without being sensitive to the amplitude of density fluctuations. We perform several statistical efficiency and consistency tests on a mock galaxy catalogue, using the SDSS-III survey as template, taking into account the survey geometry and selection effects, to validate the Bayesian inference machinery implemented.}

   \keywords{methods: data analysis -- methods: statistical -- galaxies: statistics -- cosmology: observations -- large-scale structure of Universe}
               
    \titlerunning{Cosmological inference from Bayesian forward modelling of deep galaxy redshift surveys}
	\authorrunning{Kodi Ramanah et al.}

   \maketitle
%
%________________________________________________________________

\section{Introduction}
\label{intro}

The past few decades have witnessed the advent of an array of galaxy redshift surveys, with the state-of-the-art catalogues mapping millions of galaxies with precision positioning and accurate redshifts. The Sloan Digital Sky Survey (SDSS) \citep{sdss2000technical, sdss2009dr7, sdss2014dr10, sdss2015dr12} and the Six Degree Field Galaxy Redshift Survey (6dFGRS) \citep{6df2009dr3} are two notable examples. Future cutting-edge surveys from the Euclid \citep{euclid2011report, euclid2016missiondesign, euclid2016cosmology} and Large Synoptic Survey Telescope (LSST) \citep{lsst2008summary} missions, currently under construction, further highlight the wealth of galaxy redshift data sets which would be available within a five to ten year time frame. Sophisticated and optimal data analysis techniques, in particular large-scale structure analysis methods, are in increasing demand to cope with the present and upcoming avalanches of cosmological and astrophysical data, and therefore optimize the scientific returns of the missions.  

With the metamorphosis of cosmology into a precision (and data-driven) science, the three-dimensional (3D) large-scale structures have emerged as an essential probe of the dynamics of structure formation and evolution to further our understanding of the Universe. The two-point statistics of the 3D matter distribution have developed into key tools to investigate various cosmological models and test different inflationary scenarios. Various techniques to measure the power spectrum and several reconstruction methods attempting to recover the underlying density field from galaxy observations are described in literature \citep[e.g.][]{bertschinger1989recovering, bertschinger1991mapping, hoffman1994wiener, lahav1994wiener, fisher1995wiener, sheth1995constrained, webster1997wiener, bistolas1998nonlinear, schmoldt1999density, saunders2000interpolation, zaroubi1999wiener, zaroubi2002unbiased, erdogdu20042df, erdogdu2006reconstructed}, with the recent focus being on large-scale Bayesian inference methods \citep[e.g.][]{kitaura2008bayesian, kitaura2009cosmic, jasche2010fast, jasche2010bayesian, jasche2012bayesian, jasche2013methods, jasche2015matrix, jasche2018physical}. A formal and rigorous Bayesian framework provides the ideal setting to solve the ill-posed problem of inferring signals from noisy observations, while quantifying the corresponding statistical uncertainties.

The potential of such Bayesian algorithms to jointly infer cosmological constraints, nevertheless, has not yet been exploited. We present, for the first time, a non-linear Bayesian inference framework for cosmological parameter inference from galaxy redshift surveys via an implementation of the Alcock-Paczy\'nski \citep[AP, ][]{alcock1979evolution} test. We extend the hierarchical Bayesian inference machinery of \textsc{borg} (Bayesian Origin Reconstruction from Galaxies) \citep{jasche2013bayesian}, originally developed for the non-linear reconstruction of large-scale structures, to constrain cosmological parameters. \textsc{borg} encodes a physical model for gravitational structure formation, yielding a highly non-trivial Bayesian inverse problem. This consequently allows us to reformulate the standard problem of present 3D density field reconstruction as an inference problem for initial conditions at an earlier epoch from current galaxy observations. \textsc{borg} builds upon the implementation of the Hamiltonian Monte Carlo (HMC) method \citep{neal1993probabilistic}, initially introduced in the \textsc{hades} (HAmiltonian Density Estimation and Sampling) algorithm \citep{jasche2010fast}, for efficiently sampling the high dimensional and non-linear parameter space of possible initial conditions at an earlier epoch.

In this work, the conceptual framework is to constrain the comoving-redshift coordinate transformation and therefore infer the appropriate cosmology which would result in isotropic correlations of the galaxy density field. The key aspect of this application of the AP test consequently lies in its robustness to a misspecified model and the approximations therein, yielding a near-optimal exploitation of the model predictions, without relying on its accuracy in modelling the scale dependence of the correlations of the density field. Here, we employ Lagrangian Perturbation Theory (LPT) as a physical description for the non-linear dynamics and perform a joint inference of initial conditions, and consequently the corresponding non-linearly evolved density fields and associated velocity fields, and cosmological parameters, from incomplete observations. This augmented framework with cosmological applications is designated as \textsc{altair} (ALcock-Paczy\'nski consTrAIned Reconstruction).

The paper is organized as follows. In Section \ref{ap_test}, the underlying principles of the AP test are outlined, followed by a description of the forward modelling approach and data model implemented in Section \ref{forward_modelling}. We then test the algorithm in Section \ref{results} on an artificially generated galaxy survey, with the mock generation procedure described in the preceding Section \ref{mock_generation}, by investigating its performance via statistical efficiency and consistency tests. In Section \ref{conclusion}, we summarize the main aspects of our work and discuss further possible extensions to our algorithm in order to fully exploit its potential in deriving cosmological constraints. In Appendix \ref{lpt_poisson_posterior}, we describe the LPT-Poissonian posterior implemented in this work, followed by the computation of the Jacobian of the comoving-redshift transformation in Appendix \ref{jacobian}. We provide a brief overview of the Hamiltonian sampling approach in Appendix \ref{hmc}, and follow up by deriving the required equations of motion in Appendix \ref{equations_of_motion}, with the numerical implementation outlined in Appendix \ref{numerical_implementation}. We subsequently describe how we increase the efficiency of our cosmological parameter sampler via a rotation of the parameter space in Appendix \ref{rotation_cosmo}. Finally, we outline the derivation of the adjoint gradient for a generic 3D interpolation scheme in Appendix \ref{adjoint_interpolation}.

\section{The Alcock-Paczy\'nski test}
\label{ap_test}

\begin{figure}
	\centering
		{\includegraphics[width=\hsize,clip=true]{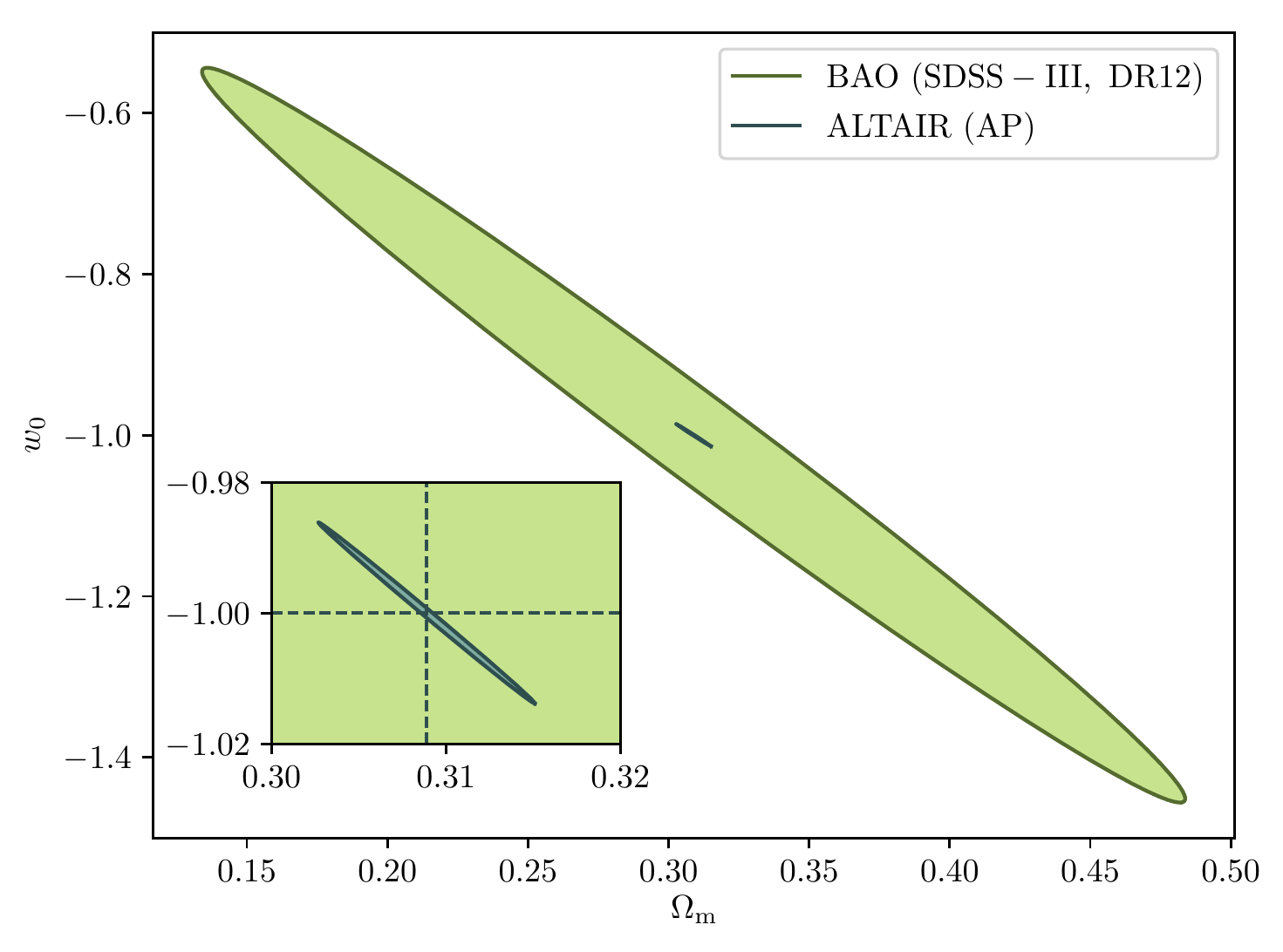}}
	\caption{Comparison of cosmological constraints from BAO measurements and our implementation of AP test in \textsc{altair}. The grey and green lines denote the $1 \sigma$ confidence region, centered on the fiducial cosmological parameters, obtained from our AP test and BAO constraints from SDSS-III (Data Release 12) \citep{alam2017clustering}, respectively. Note that the BAO constraints have not been combined with {\it Planck} CMB measurements. This demonstrates the potentially unprecedented constraining power of our AP test compared to standard BAO analyses, as discussed in Section \ref{ap_test}, with the inset focusing on the \textsc{altair} constraints where the fiducial cosmology is depicted in dashed lines. This error forecast is validated on a simulated analysis (cf. Fig. \ref{fig:marginal_posteriors_cosmo_params_altair}).}
	\label{fig:error_ellipses_BAO_altair}
\end{figure}

\begin{figure*}
	\centering
		{\includegraphics[scale=0.55]{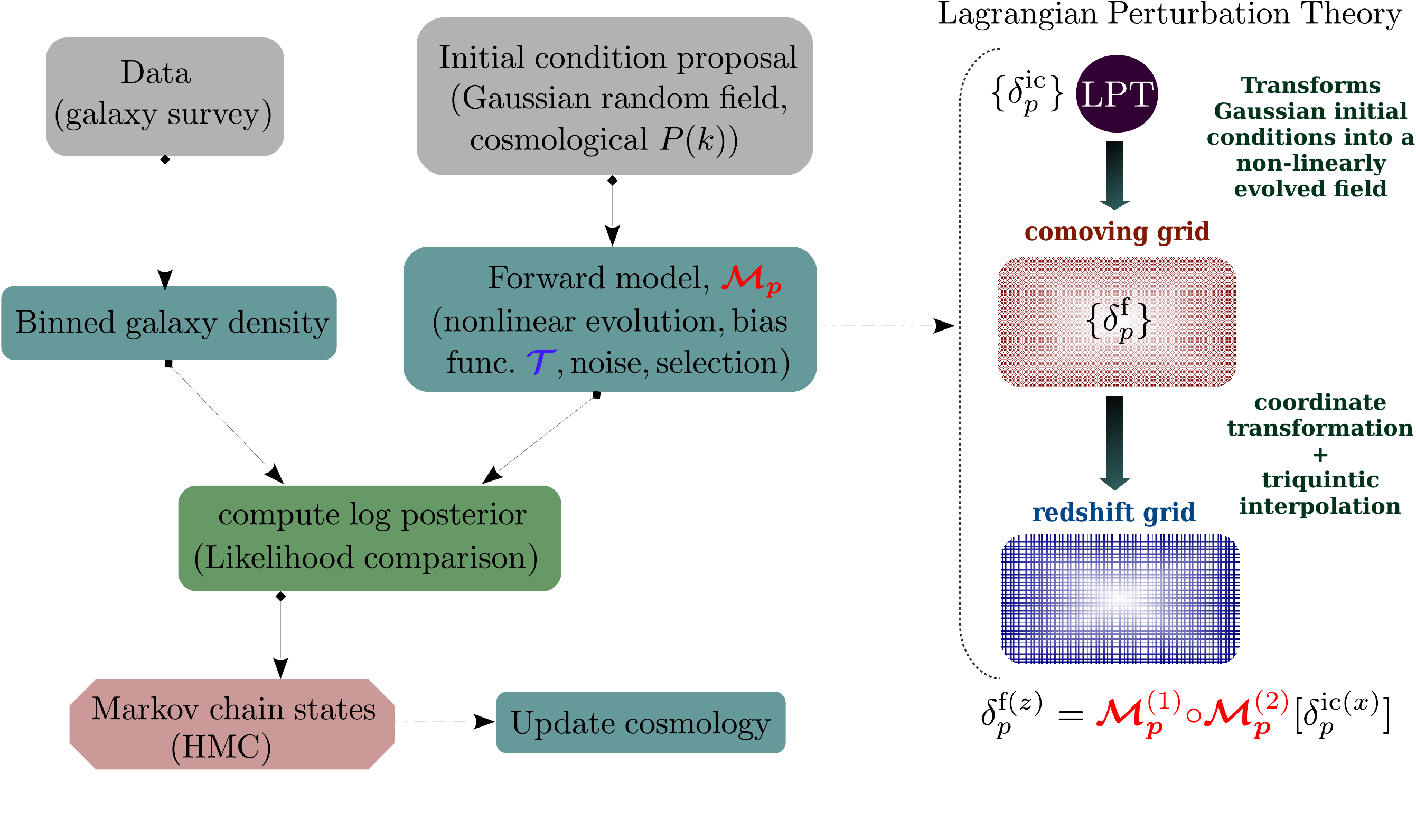}}
	\caption{Schematic representation of the reconstruction pipeline. The forward model consists of a chain of various components for the non-linear evolution from initial conditions and the subsequent transformation from comoving to redshift space for the application of the AP test. This consequently transforms the initial density field into a set of predicted observables, i.e. a galaxy distribution in redshift space, for comparison with data via a likelihood or posterior analysis.}
	\label{fig:reconstruction_schematic}
\end{figure*}

The Alcock-Paczy\'nski (AP) test \citep{alcock1979evolution} is a cosmological test of the expansion of the Universe and its geometry. The main advantage of this test is that it is independent of the evolution of galaxies but depends only on the geometry of the Universe. The assumption of incorrect cosmological parameters in data analysis  produces distortions in the appearance of any spherical object or isotropic statistical distribution. The AP test provides a pathway to exploit this resulting spurious anisotropy to constrain the cosmological parameters. Here, we invoke the AP test to ensure that the underlying geometrical properties of isotropy of the Universe \citep{friedmann1922uber, friedmann1924uber, lemaitre1927univers, lemaitre1931expansion, lemaitre1933univers, robertson1935kinematicsI, robertson1936kinematicsII, robertson1936kinematicsIII, walker1937milne, saadeh2016isotropic} are maintained. As such, the key underlying assumption adopted in this work relies purely on the geometrical symmetries of the cosmological principle. As a result, such a test does not employ the growth of structures to constrain cosmology, unlike cluster abundance \citep[e.g.][]{wang1998cluster}.

The AP test, and various formulations thereof, have been studied extensively in the context of galaxy and quasar surveys \citep[e.g.][]{phillipps1994possible, ryden1995measuring, ballinger1996measuring, matsubara1996cosmological, popowski1998quasar, delaix1998sensitivity, lopez2014alcock, li2014cosmological, li2016cosmological}. Variants of the AP test have also been successfully applied to cosmic voids \citep[e.g.][]{sutter2012first, lavaux2012precision, sutter2014measurement, hamaus2014testing, hamaus2015probing, hamaus2016constraints} and also to other cosmological observables like supernov\ae{} \citep{blake2011wigglez}, the Lyman-$\alpha$ forest \citep{hui1999geometrical} and $21$ centimetre emission maps \citep{nusser2005alcock, barkana2006separating}.

With baryon acoustic oscillations (BAOs) being a robust standard ruler, the AP test has been utilized for the simultaneous measurement of the Hubble parameter and angular diameter distance of distant galaxies \citep[e.g.][]{seo2003probing, blake2003probing, glazebrook2005measuring, padmanabhan2008constraining, shoji2009extracting}. In Fig. \ref{fig:error_ellipses_BAO_altair}, we depict the $1 \sigma$ confidence region of the cosmological constraints inferred via our implementation of the AP test. As a comparison, we also indicate the corresponding confidence region obtained via BAO measurements from the SDSS-III (Data Release 12) \citep{alam2017clustering}. These BAO constraints have not been combined with {\it Planck} measurements, which would significantly tighten the constraints. Nevertheless, this highlights the significant potential constraining power of our AP test compared to standard BAO analyses, while being at least as robust. While this improvement is extremely substantial for the mock SDSS-III survey considered here, we will investigate to what extent the above promise holds when applied to actual SDSS-III data in a follow-up work, as unknown systematics represent a potential caveat. We discuss Fig. \ref{fig:error_ellipses_BAO_altair} in more depth in Section \ref{results}.

The crucial aspect of our AP test is that it does not assume that the correlation function is correctly modelled. This robustness to a misspecified model is illustrated explicitly in Section \ref{results}, where we demonstrate that the shape of the prior power spectrum adopted in the inference framework does not impact on the inferred cosmological constraints (cf. Fig. \ref{fig:marginal_posteriors_cosmo_params_diff_Pk_altair}). As a result of this robustness, our AP test has a definite edge over standard approaches. Moreover, it has been pointed out that other cosmological tests, such as the luminosity distance - redshift relation, can be considered as generalized formulations of the AP test \citep{mukherjee2018beyond}, further underlining the strength of the approach presented in this work.

\section{The forward modelling approach}
\label{forward_modelling}

The large-scale structure (LSS) posterior implemented in this work, based on the \textsc{borg} framework \citep{jasche2013bayesian}, is described in depth in Appendix \ref{lpt_poisson_posterior}. A key component of the inference framework is the forward model $\mathcal{M}_p$ which links the initial conditions $\delta_p^{\mathrm{ic},(r)}$ to the redshift space representation of the evolved density field $\delta_p^{\mathrm{f},(z)}$ as follows:
\begin{align*}
	\delta_p^{\mathrm{f} (z)} &= \mathcal{M}_p \left( 1 + \delta_p^{\mathrm{ic} (r)} \right)
    = \mathcal{M}^{(1)}_p \circ \mathcal{M}^{(2)}_p \left( 1 + \delta_p^{\mathrm{ic} (r)} \right) = \mathcal{M}^{(1)}_p \left( \rho_p^{\mathrm{f} (r)} \right) \\
    & = \mathcal{J}_p \left( \sum_{i,j} \mathcal{E}_{ij}^{-1} \rho^{\mathrm{f} (r)}_j x^{\alpha (i)}_p y^{\beta (i)}_p z^{\gamma (i)}_p \right) - 1 , \numberthis
	\label{eq:detailed_forward_model_altair}
\end{align*}
where $\rho_p^{\mathrm{f} (r)} \equiv 1 + \delta_p^{\mathrm{f} (r)}$ is the final density field in comoving space. The forward model consists of two components, $\mathcal{M}_p = \mathcal{M}_p^{(1)} \circ \mathcal{M}_p^{(2)}$. The first component, $\mathcal{M}_p^{(2)} \equiv \mathcal{G}_p (a, \{ \delta_p^{\mathrm{ic}} \})$, contains a physical description of the non-linear dynamics, and consequently propagates the initial conditions forward in time using LPT, yielding a non-linearly evolved final density field in comoving space, $\delta^{\mathrm{f}(r)}_p$.

To encode the AP test, we incorporate another component in the forward model that takes care of the coordinate transformation from comoving $(\myvec{r})$ to redshift $(\myvec{z})$ space, encoded in $\mathcal{M}_p^{(1)}$ (cf. Fig. \ref{fig:reconstruction_schematic}). Schematically, we construct a second grid in redshift space, which involves a triquintic interpolation (fifth order interpolation scheme in three dimensions) on the comoving grid. This interpolation scheme is described in Appendix \ref{adjoint_interpolation}, with the notation employed in Eq.~\eqref{eq:detailed_forward_model_altair} clearly laid out. The corresponding Jacobian factor of this transformation, $|\mathcal{J}^z_r|$ (cf. Appendix \ref{jacobian}), entails cosmological dependence and is consequently included in the AP test as well as through the direct coordinate dependence $\mathcal{E}_{ij}$.

The redshift space representation then allows for comparison with data via the likelihood or posterior. The essence of this AP test to constrain cosmological parameters can be summarized as follows: The Bayesian inference machinery explores the various cosmological expansion histories and selects the cosmology-dependent evolution pathways which result in isotropic correlations of the galaxy density field.

Fig. \ref{fig:reconstruction_schematic} illustrates the reconstruction scheme implemented in \textsc{altair}. First, galaxies are projected from the survey onto a 3D grid, such that we have a distribution of galaxies in redshift space and this constitutes our observable. We then generate a 3D density field according to Gaussian initial conditions (homogeneous prior) with a reference power spectrum, typically $\Lambda$CDM cosmology. The forward model subsequently transforms the initial density field into a set of predicted observables which are then compared to data via a likelihood or posterior analysis. And conversely, given the position of galaxies, we can infer this density field.

While we implement LPT to approximately describe gravitational non-linear structure formation in this work, other more adequate physical descriptions such as 2LPT or the non-perturbative particle mesh \citep[see recent upgrade of \textsc{borg} in][]{jasche2018physical} can be straightforwardly employed, within the flexible block sampling approach described in Appendix \ref{block_sampling}, by upgrading the first component, $\mathcal{M}_p^{(2)} \equiv \mathcal{G}_p (a, \{ \delta_p^{\mathrm{ic}} \})$, of our forward model (cf. Fig. \ref{fig:reconstruction_schematic}). Nevertheless, our implementation of the AP test exploits essentially the isotropy of the correlation function, such that there is no explicit dependence on the accuracy of modelling the scale dependence of the correlations, rendering this method robust to a misspecified model and the approximations therein.

\subsection{The galaxy data model}
\label{poisson_likelihood}

Galaxies can be considered as tracers of the matter fluctuations since they follow the gravitational potential of the underlying matter distribution, with the statistical uncertainty due to the discrete nature of the galaxy distribution usually modelled by a Poissonian distribution \citep[e.g.][]{layser1956new, peebles1980large}. Poissonian likelihoods have emerged as the standard for non-linear LSS inference \citep[e.g.][]{jasche2010bayesian_nonlinear, jasche2010fast, kitaura2010recovering}. The Poissonian likelihood distribution implemented in \textsc{altair}, for multiple subcatalogues or galaxy observations labelled by the index $g$, can be expressed as follows:
\begin{equation}
	\mathcal{L}\left( \{ N^g_p \} | \lambda^g_p \right) = \prod_p \frac{(\lambda^g_p)^{N^g_p} \mathrm{e}^{-\lambda^g_p}}{N^g_p !} ,
	\label{eq:poissonian_likelihood_altair}
\end{equation}
where $N^g_p$ is the observed galaxy number counts in redshift space in the given voxel $p$. $\lambda^g_p$ is the expected number of galaxies at this given position and is related to the final density field $\delta_p^{\mathrm{f}}$, in redshift space, via
\begin{equation}
	\lambda^g_p \left( \{ \delta_p^{\mathrm{f}} \}, \{ \theta_i \} \right) = R^g_p (\theta_i) \mathcal{T} \left[ 1 + \delta^{\mathrm{f}}_p \right] ,
    \label{eq:mean_galaxy_number_lambda_altair}
\end{equation}
where $R^g_p$ is the overall linear response operator of the survey that incorporates the survey geometry and selection effects, and $\theta_i$ corresponds to a set of cosmological parameters. $\mathcal{T}$ is the galaxy biasing model which accounts for the fact that galaxies do not trace exactly the underlying matter distribution, and are therefore biased tracers with clustering properties that do not exactly mirror those of dark matter \citep{kaiser1984spatial}. This is currently one of the most challenging and unresolved issues hindering the analysis of galaxy distributions in non-linear regimes \citep[e.g. see the review by][]{desjacques2016largescale, schmidt2018rigorous}. 

In this work, we adopt the standard approach of a local, but non-linear bias function, in particular, the phenomenological model proposed by \cite{neyrinck2014halo}, such that the above Poisson intensity field can be expressed as
\begin{equation}
	\lambda^g_p \left( \{ \delta_p^{\mathrm{f}} \}, \{ \theta_i \}, \{ \bar{N}^g \}, \{ b^g_i \} \right) = R^g_p \bar{N}^g \left[ 1 + \delta^{\mathrm{f}}_p \right]^{\beta} \mathrm{e}^{- \rho^g [1 + \delta^{\mathrm{f}}_p]^{-\epsilon^g}} ,
	\label{eq:poisson_intensity_biased_altair}
\end{equation}
where the bias function, described by four parameters, $\bar{N}^g$, the mean density of tracers, and $\{ b^g_i \} = \{ \beta, \rho^g, \epsilon^g \}$, is a truncated power law bias model with the additional exponential function suppressing galaxy clustering in under dense regions. This bias model, with a power law and an exponential at low densities, were found to be in good agreement with standard excursion set and local-growth-factor models \citep[for more details, see][]{neyrinck2014halo}. The main limitation of this bias model is that it is purely local. Nevertheless, it is more adequate than a simplistic linear bias model and mitigates in practice the deficiencies of our physical model (LPT) at the considered resolution. The expected number of galaxies can subsequently be related to the initial conditions $\delta_p^{\mathrm{ic}}$ via the forward model $\mathcal{M}_p$, as described above, due to the deterministic nature of structure formation, i.e. the Dirac delta function in Eq.~\eqref{eq:conditional_posterior_final_density_altair}.

The logarithm of the likelihood from Eq.~\eqref{eq:poissonian_likelihood_altair} can therefore be expressed, in terms of the initial conditions, as
\begin{multline}
	\ln \mathcal{L}\left[ \{ N^g_p \} \big| \mathcal{M}_p \left(\{ \delta_p^{\mathrm{ic}} \}\right), \{ \theta_i \}, \{ \bar{N}^g \}, \{ b^g_i \} \right]  \\ \; \; \; \; \; \; \; \; \; \; \; \; \; \; = - \sum_p \Bigg\lbrace  \lambda^g_p \left( \{ \delta_p^{\mathrm{f}} \}, \{ \theta_i \}, \{ \bar{N}^g \}, \{ b^g_i \} \right) \\ - N^g_p \ln \left[ \lambda^g_p \left( \{ \delta_p^{\mathrm{f}} \}, \{ \theta_i \}, \{ \bar{N}^g \}, \{ b^g_i \} \right) \right] + \ln \left( {N^g_p !} \right) \Bigg\rbrace .
	\label{eq:log_poissonian_likelihood_altair}
\end{multline}
We therefore have a likelihood distribution that encodes the statistical process describing the generation of galaxy observations given a specific realization of 3D initial conditions. This data model is inherently non-linear as a result of the galaxy biasing model employed and also due to the signal dependence of Poissonian noise, which does not behave as an additive nuisance.

\subsection{The augmented joint posterior distribution}
\label{joint_posterior}

The augmented joint posterior distribution corresponds to the following:
\begin{multline}
	\mathcal{P} \left(\{ \delta_p^{\mathrm{ic}} \}, \{\bar{N}^g\}, \{b_i^g\}, \{ \theta_i \} |\{N^g_p\}, \mymat{S} \right) \\ \; \; \; \; \; \; \; \; \; \; \; \; \; \; \; \; \; \; \propto \mathcal{L}\left[\{N^g_p\}\big|\mathcal{M}_p (\{ \delta_p^{\mathrm{ic}} \}), \{ \theta_i \}, \{\bar{N}^g\}, \{b_i^g\} \right] \\ \times \Pi\left(\{ \delta_p^{\mathrm{ic}} \}| \, \mymat{S} \right) \Pi\left( \{\bar{N}^g \}, \{b_i^g\} \right) \Pi\left( \{\theta_i \}\right),
	\label{eq:joint_posterior_altair}
\end{multline}
where the $\Pi$'s correspond to the respective priors for each parameter. Hence, given our forward model $\mathcal{M}$, which incorporates sequential components of structure formation and coordinate transformation, as described above, the complex task of modelling accurate priors for the statistical behaviour of present-day matter fluctuations can be recast into a Bayesian inference problem for the initial conditions. From this joint posterior distribution, we can construct the various conditional posterior distributions for each parameter of interest. The modular statistical programming approach adopted in outlined in Appendix \ref{block_sampling}.

Since the non-linear LSS analysis has been reformulated as an initial conditions statistical inference problem, as described by the joint posterior distribution (\ref{eq:joint_posterior_altair}), this method depends solely on forward evaluations, and consequently has a definite edge over traditional approaches of initial conditions inference that require backward integration of the equations of motion or the inversion of the flow of time \citep[e.g.][]{nusser1992tracing}. The latter methods are prone to erroneous fluctuations in the initial density and velocity fields, resulting from spurious growth of decaying modes. Moreover, such schemes are hindered by survey incompleteness which requires the knowledge of the complex and, as yet, unknown multivariate probability distribution for the matter fluctuations, to render the backward integration of non-linear models physically meaningful via constrained realizations. In comparison, the forward modelling approach here conveniently accounts for survey masks and statistical uncertainties in the initial conditions, which amounts to modelling straightforward uncorrelated Gaussian processes, to generate data constrained realizations of the initial and evolved density fields.

\subsection{The cosmological parameter posterior distribution}
\label{theta_posterior}

In this work, we sample the present-day values of matter density and dark energy equation of state parameters, $\{\theta_i \}= \{ \Omega_{\mathrm{m}}, w_0 \}$, via the following conditional posterior distribution,
\begin{multline}
	\mathcal{P} \left( \{ \theta_i \} |\{N^g_p\}, \{ \delta_p^{\mathrm{ic}} \}, \{\bar{N}^g\}, \{b_i^g\}, \mymat{S} \right) \\ = \mathcal{P} \left( \{ \theta_i \} |\{N^g_p\}, \{ \delta_p^{\mathrm{ic}} \}, \{\bar{N}^g\}, \{b_i^g\} \right) ,
	\label{eq:conditional_cosmo_posterior_altair}
\end{multline}
assuming conditional independence of the cosmological power spectrum, i.e. Fourier transform of the covariance matrix $\mymat{S}$, once the density field is known. This assumption holds as we are only probing the cosmological expansion in this work, with the power spectrum anchored with a fiducial cosmology. We quantitatively demonstrate the validity of this assumption in Section \ref{results} by comparing the entropy of prior information against that of posterior information (cf. Fig. \ref{fig:prior_check_power_spectra_altair}). We defer power spectrum sampling to a future work. Applying Bayes' identity, and using the joint posterior distribution from Eq.~\eqref{eq:joint_posterior_altair}, we obtain
\begin{align*}
	\mathcal{P} \Big( \{ \theta_i \} &| \{N^g_p\}, \{ \delta_p^{\mathrm{ic}} \}, \{\bar{N}^g\}, \{b_i^g\} \Big) \\ 
    &= \frac{\mathcal{P} \left( \{ \theta_i \}, \{ \delta_p^{\mathrm{ic}} \}, \{\bar{N}^g\}, \{b_i^g\} |\{N^g_p\} \right)}{\mathcal{P} \left( \{ \delta_p^{\mathrm{ic}} \}, \{\bar{N}^g\}, \{b_i^g\} |\{N^g_p\} \right)} \\
    &= \frac{\mathcal{L}\left[\{N^g_p\}\big| \mathcal{M}_p \left(\{ \delta_p^{\mathrm{ic}} \} \right), \{ \theta_i \}, \{\bar{N}^g\}, \{b_i^g\} \right] }{\mathcal{P} \left( \{ \delta_p^{\mathrm{ic}} \}, \{\bar{N}^g\}, \{b_i^g\} |\{N^g_p\} \right)} \\
    & \; \; \; \; \; \; \; \; \; \times \Pi\left(\{ \delta_p^{\mathrm{ic}} \} | \, \mymat{S} \right) \Pi\left( \{\bar{N}^g \}, \{b_i^g\} \right) \Pi\left( \{\theta_i \}\right) \\
    &\propto \mathcal{L}\left[\{N^g_p\}\big| \mathcal{M}_p \left(\{ \delta_p^{\mathrm{ic}} \} \right), \{ \theta_i \}, \{\bar{N}^g\}, \{b_i^g\} \right] \times \Pi\left( \{\theta_i \}\right) ,
	\label{eq:derivation_cosmo_posterior_altair} \numberthis
\end{align*}
after omitting the terms without any cosmological parameter dependence.

Since this work attempts to demonstrate the capabilities of \textsc{altair} to constrain the cosmological parameters, as proof of concept, we set uniform prior distributions on $\theta_i$. To sample from the above marginal posterior distribution, we make use of a slice sampling procedure \citep[e.g.][]{neal2000slicesampling, neal2003slice}. After obtaining a realization of $ \theta_i$, we need to update the comoving-redshift coordinate transformation in the forward model.

We adopt a dynamical dark energy model, in particular, the standard Chevallier-Polarski-Linder (CPL) parameterization \citep{chevallier2001accelerating,linder2003exploring}, where the evolution of the dark energy equation of state parameter $w$ is a linear function of the scale factor $a$, as follows:
\begin{equation}
	w = w_0 + (1 - a)w_a  .
	\label{eq:CPL_parameterization_altair}
\end{equation} 
In this work, we set $w_a = 0$ and infer the present-day value $w_0$. Moreover, we impose the assumption of flatness, i.e. $\Omega_{\mathrm{k}} = 0$, such that the dark energy density is $\Omega_{\mathrm{de}} = 1 - \Omega_{\mathrm{m}}$.

Due to the correlation between $\Omega_{\mathrm{m}}$ and $w_0$, we perform a rotation of the $(\Omega_{\mathrm{m}}, w_0)$ parameter space,  using orthonormal basis transformations derived from the covariance matrix, to improve the efficiency of the slice sampler. This procedure is outlined in Appendix \ref{rotation_cosmo}, with the significant gain in efficiency illustrated in Fig. \ref{fig:correlation_length_cosmo_MCMC_chains_altair}. The corresponding sampler for the bias parameters is outlined in Appendix \ref{bias_posterior}.

\section{Generation of a mock galaxy catalogue}
\label{mock_generation}

\begin{figure*}
	\centering
		{\includegraphics[width=\hsize,clip=true]{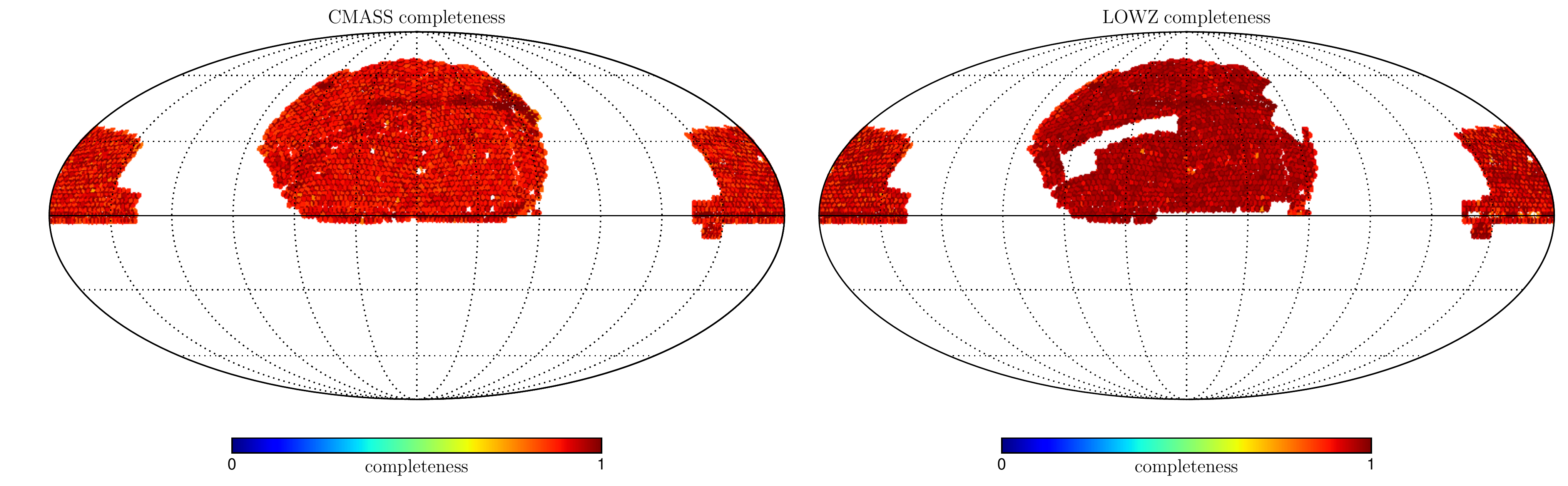}}
	\caption{The observed sky completeness, used to generate and analyze the mock catalogue in this work, are illustrated in the left and right panels, corresponding to the CMASS and LOW-Z components of the SDSS-III survey, respectively.}
    \label{fig:completeness_maps}
\end{figure*}

\begin{figure}
	\centering
		{\includegraphics[width=\hsize,clip=true]{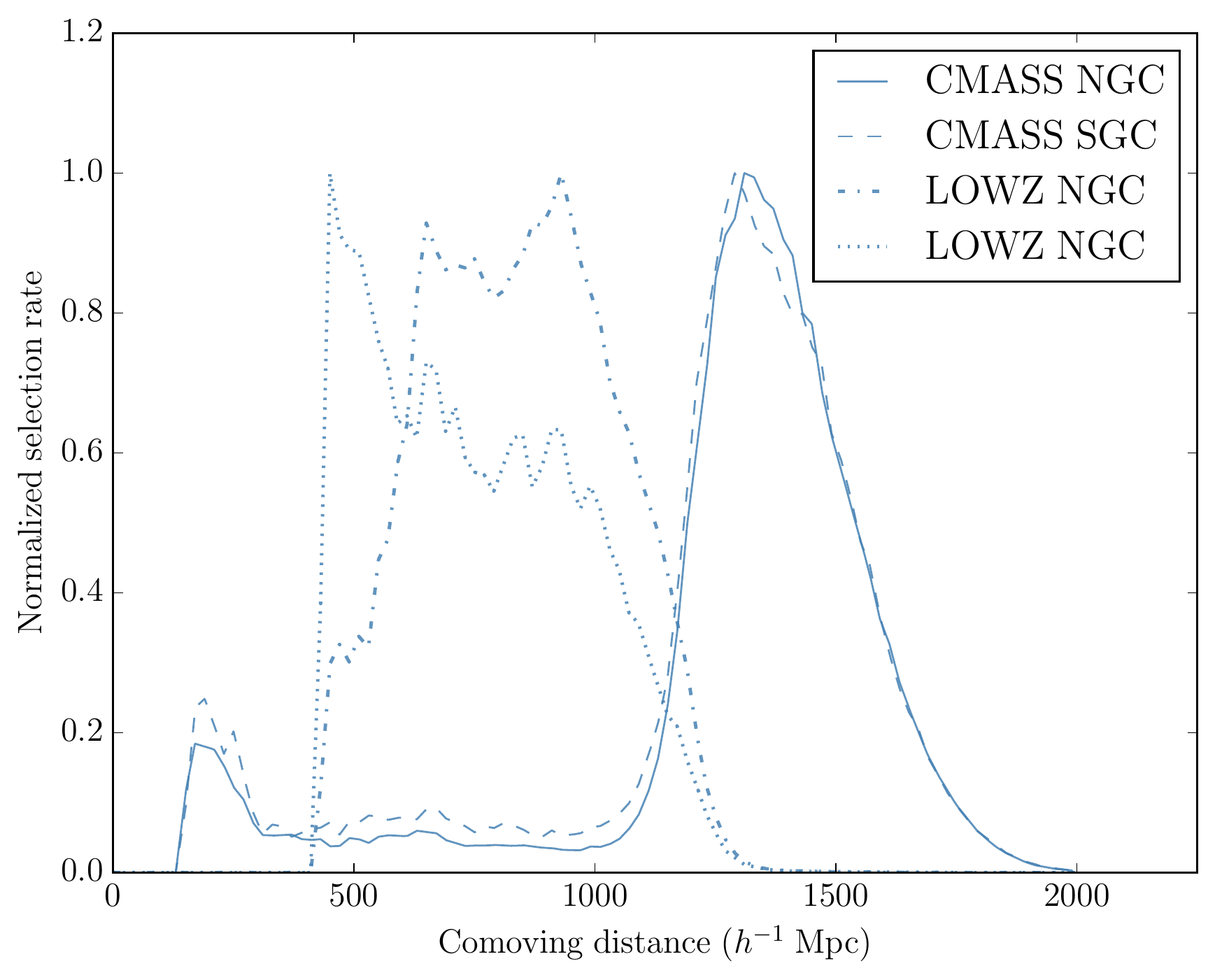}}
	\caption{The radial selection functions for the CMASS sample, in solid and dashed lines for the North Galactic Cap (NGC) and South Galactic Cap (SGC), respectively. The corresponding radial selection functions for the LOW-Z sample are depicted in dash-dotted (NGC) and dotted lines (SGC). These selection functions are used to generate the mock data to emulate features of the actual SDSS-III BOSS data.}
    \label{fig:radial_selection}
\end{figure}

%%% RESULTS PLOTS BELOW

\begin{figure*}
	\centering
		{\includegraphics[width=0.8\hsize,clip=true]{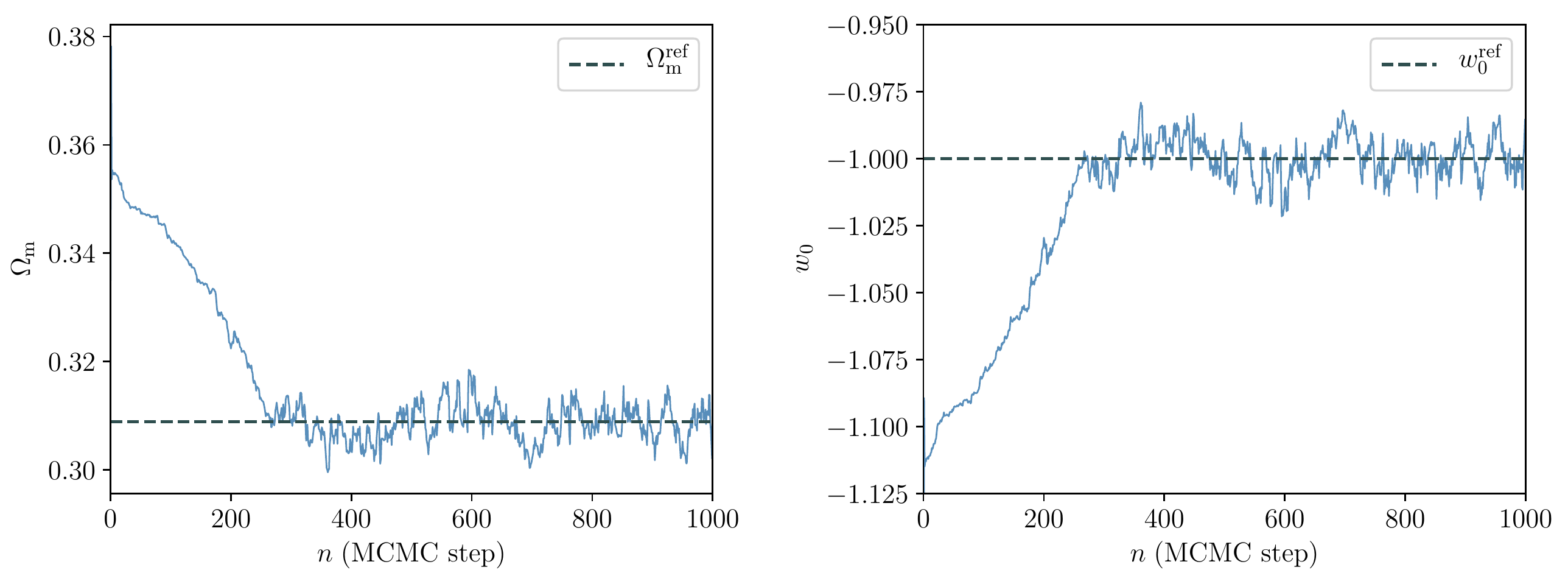}}
	\caption{The MCMC chains for the cosmological parameters, for the first 1000 samples, with the reference cosmology employed in the mock generation indicated by the horizontal dashed lines. An initial burn-in phase lasting $\sim$250 Markov transitions is illustrated by the coherent drift of the Markov chain towards the preferred region in parameter space.}
	\label{fig:cosmo_MCMC_chains_altair}
\end{figure*}

\begin{figure*}
	\centering
    {\includegraphics[width=\hsize,clip=true]{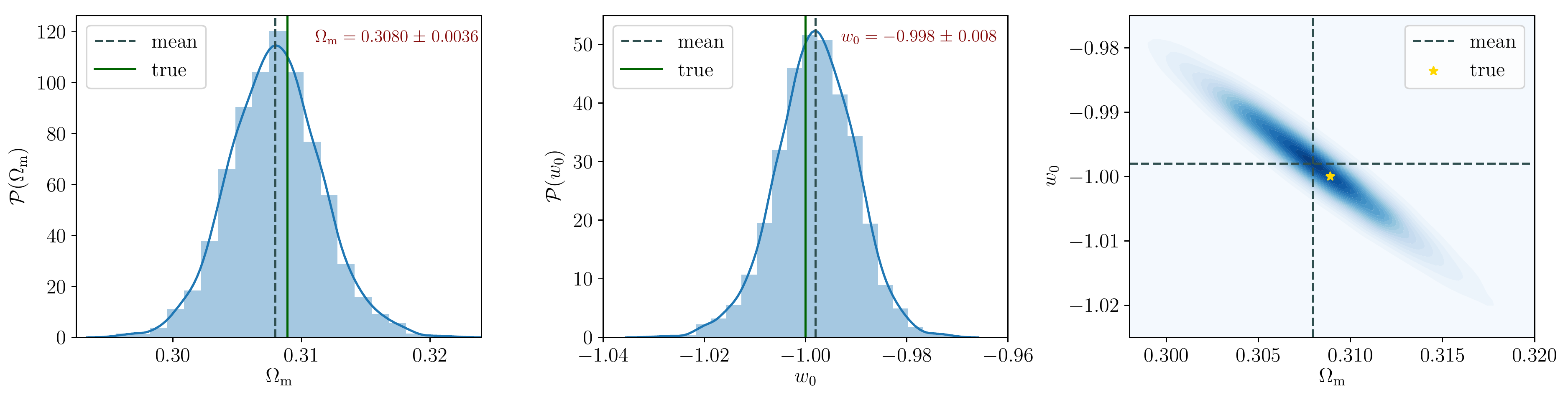}}   
	\caption{The marginal posteriors for $\Omega_{\mathrm{m}}$ ({\it left panel}) and the dark energy equation of state, $w_0$ ({\it middle panel}), for $\sim$3000 MCMC realizations, ignoring the burn-in phase of $\sim$250 Markov steps. The corresponding mean and standard deviation for each parameter are indicated in the top right corner of each plot. The joint posterior ({\it right panel}) for $\Omega_{\mathrm{m}}$ and $w_0$, depicting the high level of correlation between these two parameters. The highly informative distortion due to the cosmic expansion, as a result of probing a deep redshift range, yields extremely tight constraints on the above cosmological parameters. As a consistency test, this validates our implementation of the AP test to correctly recover the input cosmology.}
    \label{fig:marginal_posteriors_cosmo_params_altair}
\end{figure*}

We describe the generation of an artificial galaxy survey using as template the Sloan Digital Sky Survey (SDSS-III), consisting of four galaxy subcatalogues, to validate the methodology described in the previous sections. The procedure implemented here for the mock generation is essentially based on the descriptions provided in \cite{jasche2010fast} and \cite{jasche2013bayesian}. 

At first, we generate a realization for the initial density contrast $\delta_p^{\mathrm{ic}}$ from a normal distribution with zero mean and covariance corresponding to an underlying cosmological power spectrum for the matter distribution. This power spectrum, including baryonic wiggles, is computed following the prescription provided in \cite{eisenstein1998baryonic, eisenstein1999power}, assuming a standard $\Lambda$~cold dark matter ($\Lambda$CDM) cosmology with the set of cosmological parameters ($\Omega_{\mathrm{m}} = 0.3089$, $\Omega_\Lambda = 0.6911$, $\Omega_{\mathrm{b}} = 0.0486$, $h = 0.6774$, $\sigma_8 = 0.8159$, $n_{\mathrm{s}} = 0.9667$) from {\it Planck}  \citep{13planck2015}. This yields a 3D Gaussian initial density field in an equidistant Cartesian grid with $N_{\mathrm{side}} = 128$, i.e. consisting of $128^3$ voxels, where each voxel corresponds to a discretized volume element with a physical voxel size of 15.6 \Mpch, and comoving box length of $4000$ \Mpch. As a result, this implies a total of $\sim 2.1 \times 10^6$ inference parameters, corresponding to the amplitude of the primordial density fluctuations at the respective grid nodes.

To ensure a sufficient resolution of inferred final density fields, we oversample the initial density field by a factor of eight, thereby evaluating the LPT model with $256^3$ particles in every sampling step. The following step is to scale this 3D distribution of initial conditions to a cosmological scale factor of $a_{\mathrm{init}} = 0.001$ via multiplication with a cosmological growth factor $D^+ (a_{\mathrm{init}})$. These initial conditions are subsequently evolved forward in time, in non-linear fashion, with LPT providing an approximate description of gravitational LSS formation. We then construct a final 3D non-linearly evolved density field $\delta_p^{\mathrm{f}}$ from the resulting particle distribution via the cloud-in-cell (CIC) method \citep[e.g.][]{hockney1988computer}.

To generate the mock galaxy survey, we essentially need to simulate the inhomogeneous Poissonian process described by Eq.~\eqref{eq:poissonian_likelihood_altair}, by drawing random samples from the distribution, on top of the final density field $\delta_p^{\mathrm{f}}$. In this work, we generate a mock data set with realistic features emulating the highly structured survey geometry and selection effects of the SDSS-III survey, with the observed sky completeness depicted in the left and right panels of Fig. \ref{fig:completeness_maps}, respectively, for the CMASS and LOW-Z components of the SDSS-III survey. To account for the different selection effects in the northern and southern galactic planes, each component is further divided into two subcatalogues, corresponding to North/South Galactic Caps (NGC/SGC). The respective radial selection functions for these four subcatalogues are illustrated in Fig. \ref{fig:radial_selection}. Here, the selection functions are numerical estimates obtained by binning the corresponding distribution of tracers $N(d)$ in the CMASS and LOW-Z components \citep[e.g.][]{ross2017clustering}, where $d$ is the comoving distance from the observer.

The projection of the completeness functions into the 3D volume produces the 3D observation mask $C^g_p$. The survey properties are described by the galaxy selection function and the 3D completeness function, with the product of these two functions yielding the survey response operator $R^g_p$. More specifically, it is the average of the product of the 2D survey geometry $C^g(\hat{x}) = C^g_{p(\hat{x})}$ and the selection function $f(x)$ at each volume element of the 3D grid:
\begin{equation}
	R^g_p = \frac{1}{|\mathcal{V}_p|} \int_{\mathcal{V}_p} \mathrm{d}^3 \myvec{x} \; C^g(\hat{x}) f^g(x) ,
	\label{eq:survey_response_integral_altair}
\end{equation}
where the volume occupied by the $p$th voxel is indicated by $\mathcal{V}_p$, with $|\mathcal{V}|$ the volume of the set $\mathcal{V}$.

Finally, we generate the four artificial galaxy subcatalogues, labelled by $g$, by Poisson sampling on the grid with ${\bar{N}}^g_p = \{ 110.42, 122.94, 71.43, 205.48 \}$, resulting in a total number of $997828$ galaxies. The values of ${\bar{N}}^g_p$ are chosen such that the mock catalogue reflects the characteristics of the actual SDSS-III data which contains around one million tracers.

\section{Results}
\label{results}

\begin{figure*}
	\centering
		{\includegraphics[width=\hsize,clip=true]{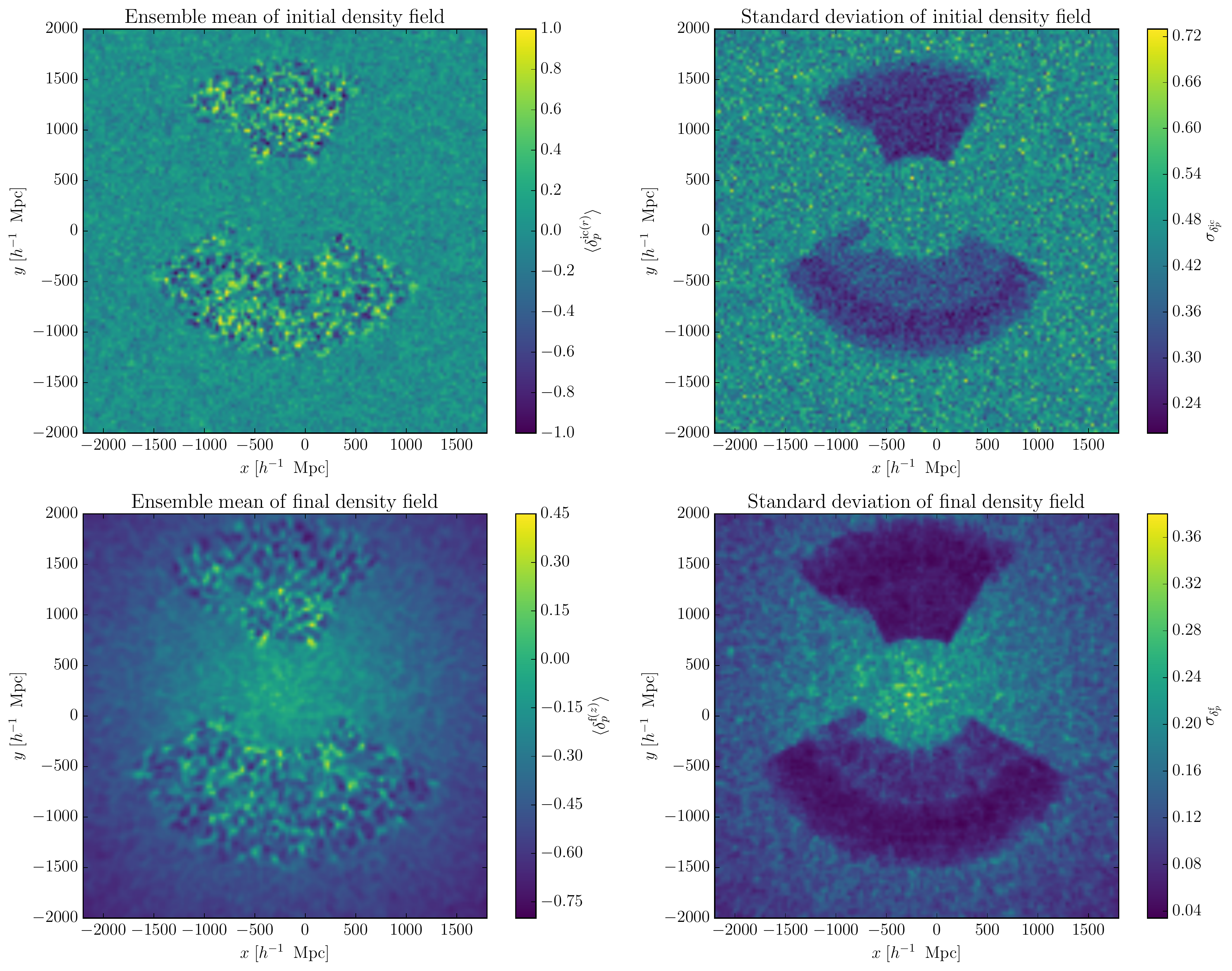}}
	\caption{The mean and standard deviation maps for the initial ({\it top panel}) and final density fields ({\it bottom panel}), computed from the MCMC realizations, with the same slice through the 3D density fields being illustrated above. In unobserved or masked regions, the density fields are not constrained by data, and they average out to the cosmic mean density. However, in observed regions, the Gaussian nature of the initial conditions and the filamentary nature of the non-linearly evolved density field is manifest. The corresponding variance is therefore higher in regions devoid of data.}
	\label{fig:mean_std_dev_density_maps_altair}
\end{figure*} 

\begin{figure}
	\centering
		{\includegraphics[width=\hsize,clip=true]{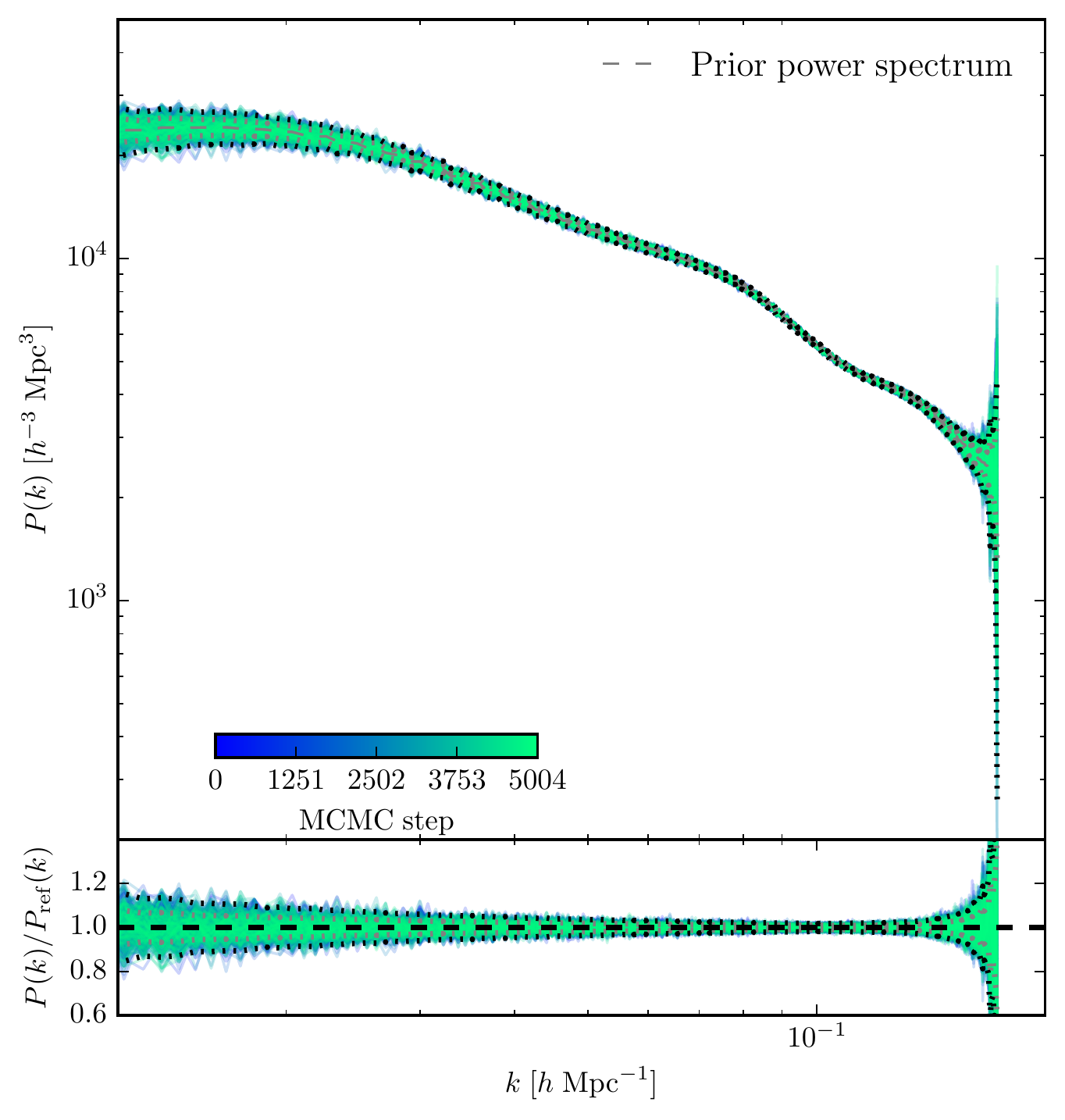}}
	\caption{{\it Top panel:} The reconstructed power spectra from the inferred posterior initial density field realizations for all sampling steps of the Markov chain. The MCMC samples are colour-coded according to their sample number, as indicated by the colour bar. This is a self-consistency test to verify whether the sampled density fields are in accordance with the reference power spectrum, depicted in dashed lines, employed in the mock generation. {\it Bottom panel:} The deviation from the true underlying solution, with the thick dashed lines representing the prior power spectrum. The thin grey and and black dotted lines correspond to the Gaussian $1\sigma$ and $2\sigma$ limits, respectively.}
	\label{fig:reconstructed_power_spectra_altair}
\end{figure}

\begin{figure}
	\centering
		{\includegraphics[width=\hsize,clip=true]{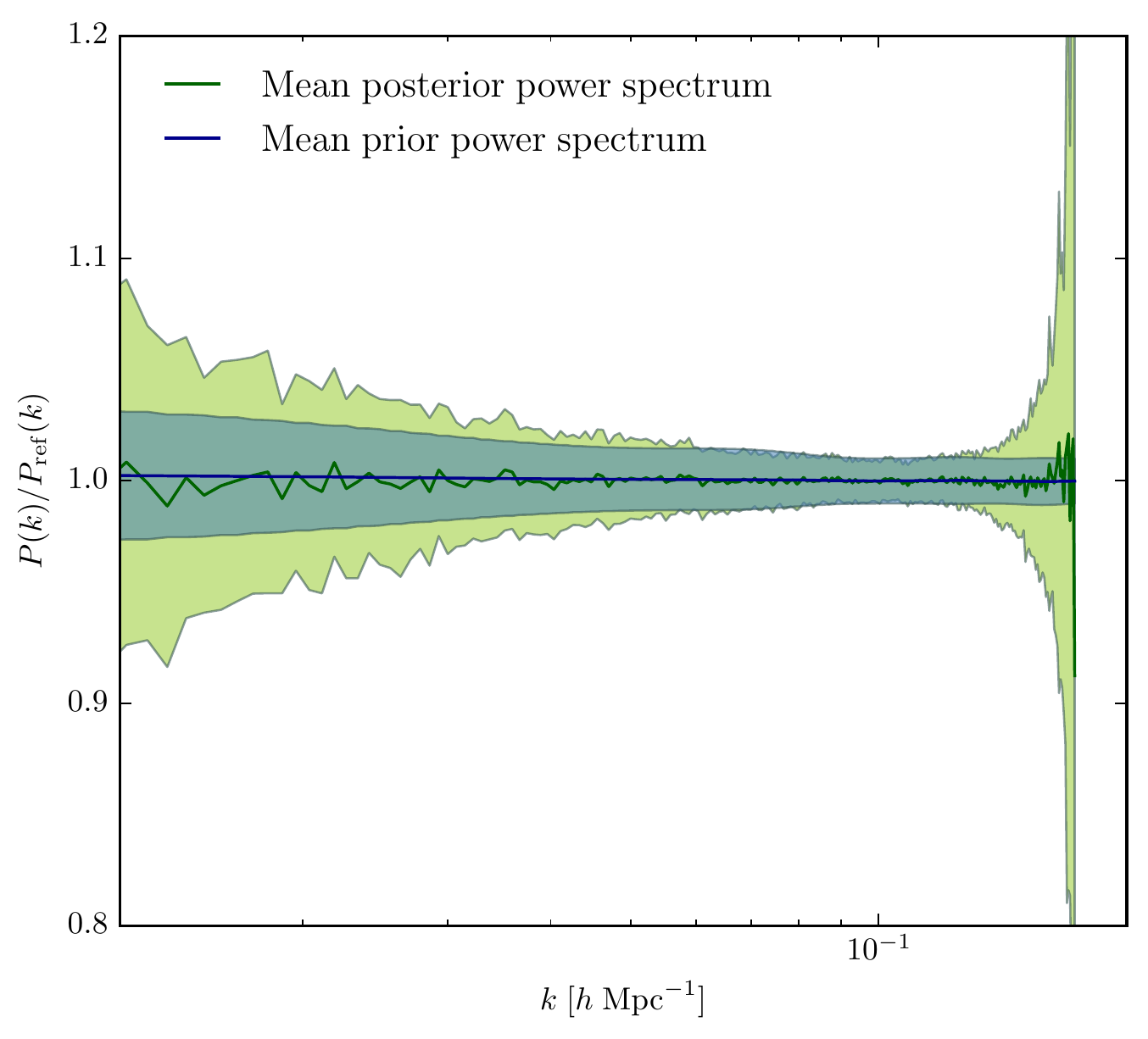}}
	\caption{The summary statistics of the reconstructed power spectra from the inferred posterior initial density field realizations depicted in Fig. \ref{fig:reconstructed_power_spectra_altair}, with the ensemble mean indicated by a solid green line. The solid blue line corresponds to the ensemble mean of the power spectra realizations generated using the inferred cosmological parameters. The shaded regions indicate their respective $1\sigma$ confidence region, i.e. 68\% probability volume. This plot shows that the prior information entropy is inferior to the posterior information entropy, due to the narrower distribution of the former. The prior power spectrum adopted, as a result, does not impact significantly on the cosmological parameter inference via the AP test.} 
	\label{fig:prior_check_power_spectra_altair}
\end{figure}

\begin{figure*}
	\centering
    {\includegraphics[width=\hsize,clip=true]{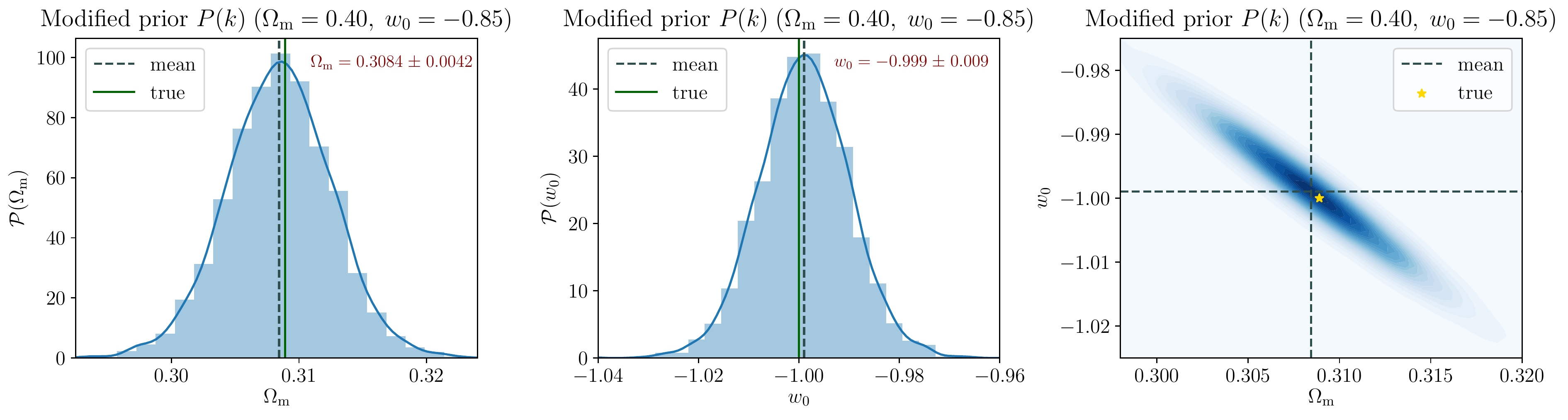} }        
	\caption{Same as Fig. \ref{fig:marginal_posteriors_cosmo_params_altair}, but employing a different prior power spectrum ($\Omega_{\mathrm{m}} = 0.40$, $w_0 = -0.85$), for $\sim$3000 MCMC realizations. By recovering the fiducial cosmological parameters employed in the mock generation, this test case explicitly highlights the robustness of our approach to the shape of the prior power spectrum adopted. The corresponding uncertainties are slightly larger than for the original run by around 15\%.}
    \label{fig:marginal_posteriors_cosmo_params_diff_Pk_altair}
\end{figure*}

\begin{figure}
	\centering
		{\includegraphics[width=\hsize,clip=true]{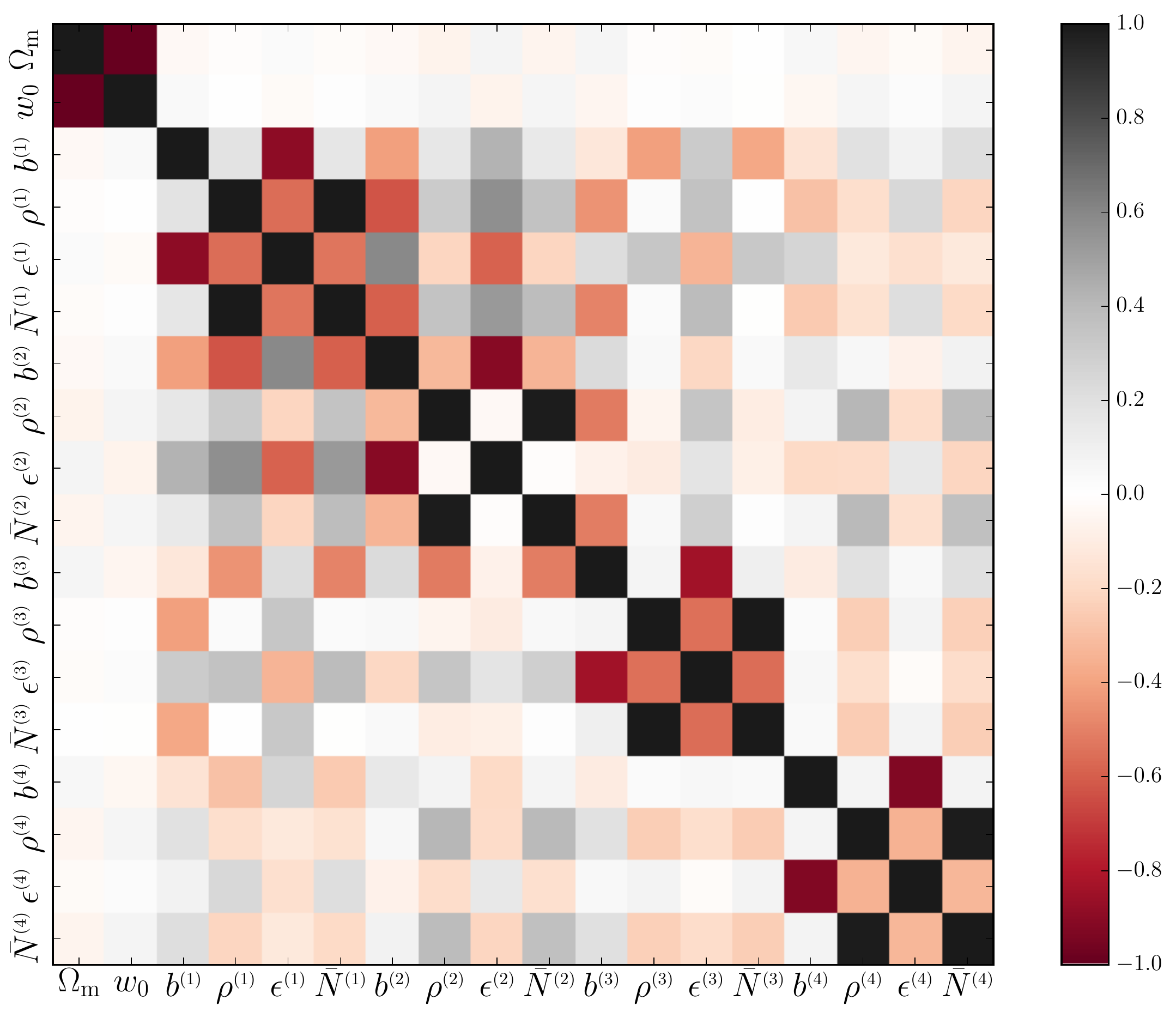}}
	\caption{Correlation matrix of the galaxy bias and cosmological parameters, normalized using their respective variance. This illustrates the weak correlation between the inferred cosmological constraints and the galaxy bias. The lack of dependence on the currently unknown phenomenon of galaxy biasing is therefore a key highlight of our implementation of the AP test for cosmological parameter inference.}
	\label{fig:cosmo_bias_params_covariance_altair}
\end{figure}

We perform a series of tests to evaluate the performance of our algorithm in a realistic context by applying \textsc{altair} on the simulated SDSS-III-like galaxy catalogue described in the previous section. In particular, the focus is on the burn-in and convergence behaviour of our method, which are key indicators of the overall numerical feasibility and statistical efficiency for real data applications. To validate the conceptual framework for cosmological parameter inference and the robustness of our implementation of the AP test, the reproducibility of the input cosmology and the correlations with the other inferred quantities such as galaxy bias are also of interest.

The Markov chains for the cosmological parameters, displayed in Fig. \ref{fig:cosmo_MCMC_chains_altair}, were initialized with an over-dispersed state. This figure consequently illustrates an initial burn-in phase, lasting $\sim$250 MCMC steps, where the Markov chains follow a persistent drift towards the high probability region of the parameter space. The rotation of the $( \Omega_{\mathrm{m}}, w_0 )$ parameter space before slice sampling, as described in Appendix \ref{rotation_cosmo}, reduces the burn-in period significantly, by roughly a factor of five, as shown in the right panel of Fig. \ref{fig:correlation_length_cosmo_MCMC_chains_altair}, resulting in improved sampling efficiency.

The corresponding marginal and joint posterior distributions for the cosmological parameters are displayed in Fig. \ref{fig:marginal_posteriors_cosmo_params_altair}, demonstrating the capability of \textsc{altair} to infer tight constraints from galaxy redshift surveys. This robust AP test fully exploits the high information content from the cosmic expansion as a result of probing a deep redshift range, where the distortion is more pronounced, yielding the following cosmological constraints: $\Omega_{\mathrm{m}} = 0.3080 \pm 0.0036$ and $w_0 = -0.998 \pm 0.008$. As a comparison, the SDSS-III (DR12, BAO $+$ {\it Planck}) constraints are as follows: $\Omega_{\mathrm{m}} = 0.310 \pm 0.005$ and $w_0 = -1.01 \pm 0.06$ \citep{alam2017clustering}, further highlighting the significant constraining power of our AP test. We acknowledge the significant difference in the size of uncertainties. A back of the envelope computation of the information gain is as follows: Considering a sphere of 100 $h^{-1}$~Mpc for BAO against all voxels in 4000 \Mpch, $N_\text{BAO}=4000^3 / (4\pi \times 100^3/3) = 15278$, compared to $N_\text{vox}=128^3$, yields an improvement of $\sqrt{N_\text{vox}/N_\text{BAO}} = 11.7$, which provides an order of magnitude of our uncertainties on the cosmological parameters. This is an approximate attempt to quantify the information gain from including smaller scales (in our work, $\sim0.17$ \Mpch) than the BAO scale by essentially counting the number of modes. However, the above argument does not imply that employing finer resolutions will result in an infinite gain of information. There is a saturation of information at a certain resolution due to the slow variation in the density fields across neighbouring voxels, such that further refinement beyond this limit will not yield any additional information.

To verify that the cosmological information stems purely from the geometric distortion due to the cosmic expansion, i.e. the AP test, we perform the following experiment: We deactivate the cosmic expansion component in our forward model and sample the cosmological parameters using only the LPT evolved density field. We consequently recover the corresponding prior distributions for the marginal posteriors of the cosmological parameters. As a result, this implies that the information derives purely from the geometry and not from the clustering of the non-linearly evolved density field, at least for the test case with the physical voxel size considered in this work.

The mean initial and final density fields computed from $\sim$4000 realizations (ignoring the burn-in phase) are illustrated in the left panel of Fig. \ref{fig:mean_std_dev_density_maps_altair} for a particular slice of the 3D field. The maps reveal that on average we can recover highly detailed and well-defined structures in the observed regions. In particular, the filamentary nature of the non-linearly evolved density field can be clearly seen, while the Gaussian nature of the initial conditions can also be deduced visually. However, in poorly or not observed regions, the ensemble mean density field approaches the cosmic mean density, as expected, since regions lacking any observational information should on average reflect the cosmic mean. The uncertainty in these regions is accurately accounted for in the inference process, as demonstrated by the right panel, since each sampled density field is a constrained realization, i.e. these regions are augmented with statistically correct information. 

In the top panel of Fig. \ref{fig:reconstructed_power_spectra_altair}, we illustrate the power spectra reconstructed from all the realizations of 3D initial density field, obtained from the posterior via the HMC sampler. This is a self-consistency test to confirm that the sampled density fields are in agreement with the reference (prior) power spectrum adopted for the mock generation, as substantiated quantitatively in the bottom panel, where the ratio of the {\it a posteriori} power spectra to the prior distribution is shown. The measured power spectra therefore demonstrate that the individual realizations possess the correct power throughout the entire domain of Fourier modes considered in this work. Moreover, they do not exhibit any spurious power artefacts typically resulting from erroneous treatment of survey characteristics, such as survey geometry and selection effects, and galaxy biases, implying that such effects have been properly accounted for.

In order to verify the impact of the prior power spectrum on the actual inference of cosmological parameters, we illustrate, in Fig. \ref{fig:prior_check_power_spectra_altair}, the distributions of power spectra computed using the inferred cosmology and the prior analytic prescription \citep{eisenstein1998baryonic,eisenstein1999power} and the reconstructed power spectra from Fig. \ref{fig:reconstructed_power_spectra_altair}, via their respective summary statistics, normalized with respect to the fiducial power spectrum. The scatter in the latter {\it a posteriori} power spectra reconstructed from the sampled density field realizations is significantly higher than distribution of the former prior spectra. This implies that the entropy of prior information is much lower than that of posterior information, thereby demonstrating that the prior power spectrum does not influence the cosmological parameter inference via the AP test and justifying the assumption made in Section \ref{theta_posterior}. 

We further demonstrate this robustness of our AP test by employing a modified prior power spectrum in the inference procedure. By adopting a different cosmology ($\Omega_{\mathrm{m}} = 0.40$, $w_0 = -0.85$), we modify the shape of the power spectrum, and subsequently apply \textsc{altair} on the same mock catalogue from Section \ref{mock_generation}. As shown in Fig. \ref{fig:marginal_posteriors_cosmo_params_diff_Pk_altair}, we recover the fiducial cosmological parameters employed in the mock generation, although with slightly larger uncertainties than for the original run by roughly 15\%. This test case therefore explicitly highlights the robustness of our implementation of the AP test to a misspecified model since it does not optimize the information from the scale dependence of the correlations of the density field, but rather from the isotropy of the field.

From the correlation matrix of $\Omega_{\mathrm{m}}$, $w_0$ and the galaxy bias parameters for each of the four subcatalogues, illustrated in Fig. \ref{fig:cosmo_bias_params_covariance_altair}, we deduce the extremely weak correlation between the cosmological constraints and the bias. This is a key positive aspect of our method, as galaxy bias remains nevertheless a highly active and challenging field of research \citep[see, for e.g.,][]{desjacques2016largescale}, due to its complex non-linear behaviour on intermediate and small scales, which may potentially limit the effectiveness of traditional methods of cosmological parameter inference \citep{pollina2018relative}.

Moreover, this insensitivity to the galaxy bias implies that our method does not rely on absolute density fluctuation amplitudes, but on the actual location of matter. This entails that our AP test exploits the geometrical structure of the density field and not its absolute amplitude since the power spectrum does not influence the inferred cosmological constraints. To the best of our knowledge, this is a novel aspect of cosmological inference, with most of our current understanding of cosmology based on measurements of density contrast amplitudes. We present, therefore, one of the first methods which extracts a large fraction of information from statistics other than that of direct density contrast correlations, without relying on the power spectrum or bispectrum. Our method consequently yields complementary information to state-of-the-art methods. 

\section{Summary and conclusions}
\label{conclusion}

We presented the implementation of a robust AP test that performs a detailed fit of the cosmological expansion via a non-linear and hierarchical Bayesian LSS inference framework. This forward modelling approach employs LPT as a physical description for the non-linear dynamics and sequentially encodes the cosmic expansion effect for joint inference of the cosmological parameters and underlying 3D density fields, while also fitting the mean density of tracers and bias parameters. In essence, this inference machinery explores the various cosmological expansion histories and selects the cosmology-dependent evolution pathways which yield isotropic correlations of the galaxy density field, thereby constraining cosmology.

We demonstrated the application of our algorithm \textsc{altair} on an artificially generated galaxy catalogue, consisting of four subcatalogues, that emulates the highly structured survey geometry and selection effects of SDSS-III. We performed a series of statistical efficiency and consistency tests to validate the methodology adopted and showcased its potential to yield tight constraints on cosmological parameters from current and future galaxy redshift surveys. The main strength of our implementation of the AP test lies in its robustness to a misspecified model and its inherent approximations, thereby near-optimally exploiting the model predictions, without relying on its accuracy in modelling the scale dependence of the correlations of the field.

Moreover, another key aspect of our approach, resulting from the robustness to a misspecified model, is that the cosmological constraints show extremely weak dependence on galaxy bias. This yields two crucial advantages. First, this is especially interesting as the lack of a sufficient physical description of this bias remains a potential limiting factor for standard approaches of cosmological parameter inference from such redshift surveys. Furthermore, this lack of sensitivity to the bias also implies that our method does not depend on the absolute density fluctuation amplitudes. This is therefore among the first methods to extract a large amount of information from statistics other than that of direct density contrast correlations, without relying on the power spectrum or bispectrum, thereby providing complementary information to state-of-the-art techniques.

There is scope for further development of the \textsc{altair} framework, such as incorporating power spectrum inference, which is a highly non-trivial undertaking. We also intend to augment the current formalism to include the treatment of redshift space distortions, which is key for unbiased constraints on the cosmological parameters, and apply \textsc{altair} on state-of-the-art galaxy redshift catalogues for cosmological inference.

\section*{Acknowledgements}

We express our appreciation to the anonymous reviewer for their constructive comments that helped us to improve the manuscript. We are grateful to Franz Elsner, Florent Leclercq and Natalia Porqueres for interesting discussions and/or for their comments on a draft version of the paper. We thank Tom Charnock for his help with coining the name \textsc{altair} in a coherent context. We also thank Jeremy Tinker for his assistance in interpreting the SDSS-III (DR12) science products. We acknowledge financial support from the ILP LABEX (under reference ANR-10-LABX-63) which is financed by French state funds managed by the ANR within the Investissements d'Avenir programme under reference ANR-11-IDEX-0004-02.  This work was supported by the ANR BIG4 project, grant ANR-16-CE23-0002 of the French Agence Nationale de la Recherche. BDW is supported by the Simons Foundation. This work has made use of the {\it Horizon/Beyond} cluster at the Institut d'Astrophysique de Paris. This work is supported, through  the maintenance of {\it Horizon/Beyond} cluster, under the name ORIGIN by the Domaine d’Intérêt Majeur (DIM) Astrophysique et Conditions d’Apparition de la Vie (ACAV), and received financial support from Région Ile-de-France. We thank St\'ephane Rouberol for running smoothly this cluster for us. This work is done within the Aquila Consortium.\footnote{\url{https://aquila-consortium.org}}

\bibliographystyle{aa.bst}
\bibliography{compiled_references}

\appendix
%\onecolumn 

\section{The LPT-Poissonian posterior}
\label{lpt_poisson_posterior}

In this section, we describe the large-scale structure (LSS) posterior distribution implemented in this work. We demonstrate how the complex problem of exploring the high dimensional joint posterior distribution can be reduced to a set of distinct steps via a multiple block sampling scheme.

\subsection{The density posterior distribution}
\label{density_posterior}

The primary objective is to fully characterize the 3D cosmic LSS in observations via a numerical representation of the corresponding LSS posterior using sophisticated Markov Chain Monte Carlo (MCMC) techniques, in particular to provide data constrained realizations of a set of plausible 3D density contrast amplitudes underlying a given set of galaxy observations. The posterior distribution for the evolved density field fluctuation $\delta_p^{\mathrm{f}}$ can be formulated, in a general context, via Bayes' identity as:
\begin{equation}
	\mathcal{P} \left( \{ \delta_p^{\mathrm{f}} \} | \{ N^g_p \} \right) = \frac{\mathcal{L}\left( \{ N^g_p \} | \{ \delta_p^{\mathrm{f}} \} \right) \Pi\left( \{ \delta_p^{\mathrm{f}} \} \right)}{\Pi\left( \{ N^g_p \} \right)}  ,
	\label{eq:bayes_density_posterior_altair}
\end{equation} 
where $N^g_p$ is the observed number of galaxies in voxel $p$, at position $\myvec{x}_p$ in the sky, in redshift space, for the $g$th galaxy sample, with the prior $\Pi (\{ \delta_p^{\mathrm{f}} \})$ incorporating our {\it a priori} knowledge of the present-day matter fluctuations in the Universe, the likelihood $\mathcal{L}( \{ N^g_p \} | \{ \delta_p^{\mathrm{f}} \} )$ and the normalizing factor given by the evidence $\Pi( \{ N^g_p \} )$. 

A major stumbling block consequently arises, as discussed extensively in \cite{jasche2013bayesian}, since the inference framework requires a suitable prior $\Pi (\{ \delta_p^{\mathrm{f}} \})$ which adequately describes the physical behaviour of the gravitationally evolved density field. Nevertheless, as elaborated in the following section, most attempts made in this direction so far have been based on heuristic approximations and the absence of a closed form description of the present day matter fluctuations encoded in a multivariate probability density distribution still persists.

However, \cite{jasche2013bayesian} proposed an elegant approach to circumvent this key impediment based on the following assertions: There is substantial evidence that primordial seed fluctuations at redshifts $z \sim 1000$ can be modelled as a Gaussian random field to great accuracy \citep[e.g.][]{linde2008inflationary, komatsu2011sevenyear, planck2016nongaussianity}, consistent with inflationary theories and CMB observations. Moreover, the evolution of the initial conditions relies solely on deterministic gravitational structure formation processes. Therefore, a conceptually reasonable alternative to modelling the complex statistical behaviour of the non-linear matter distribution is to formulate the inference problem at the level of the initial conditions adequately described by Gaussian statistics. This constitutes the conceptual foundation of the \textsc{borg} framework \citep{jasche2013bayesian}.

Given a forward model $\mathcal{M}_p$ that connects the initial conditions $\delta^{\mathrm{ic}, (r)}_p$ , in comoving $(\myvec{r})$ space, to the redshift $(\myvec{z})$ space representation of the final density field $\delta^{\mathrm{f},(z)}_p$, we can therefore express the conditional posterior for the evolved density field as 
\begin{equation}
	\mathcal{P}\left( \{ \delta^{\mathrm{f}}_p \} | \{ \delta_p^{\mathrm{ic}} \} \right) = \prod_p \delta^{\mathrm{D}} \left[ \delta^{\mathrm{f}}_p - \mathcal{M}_p \left( \{ \delta_p^{\mathrm{ic}} \} \right) \right] ,
	\label{eq:conditional_posterior_final_density_altair}
\end{equation}
where $\delta^{\mathrm{D}}(x)$ denoting the Dirac delta distribution encapsulates the assumption that the structure formation process is deterministic. Within this generic framework, the forward model may be generalized to a chain of arbitrary components linking the primordial density fluctuations to the present-day density contrast. Nevertheless, at its crux lies a cosmic structure formation model $\mathcal{G}_p (a, \{ \delta_p^{\mathrm{ic}} \} )$ for the non-linear evolution of initial conditions into a final density field at a given scale factor $a$, i.e. $\delta^{\mathrm{f},(r)}_p = \mathcal{G}_p (a,\{\delta_p^{\mathrm{ic},(r)} \})$. The forward model implemented in this work to encode the AP test is described in Section \ref{forward_modelling} above.
 
We can then obtain a prior distribution for $\delta^{\mathrm{f}}_p$ via a two-step sampling procedure: First, a realization of $\delta^{\mathrm{ic}}_p$ is generated from the prior distribution $\Pi( \{ \delta_p^{\mathrm{ic}} \})$ and subsequently evolved with a given forward model $\mathcal{M}_p ( \{ \delta_p^{\mathrm{ic}} \})$. This essentially implies generating samples from the joint prior distribution of $\delta^{\mathrm{ic}}_p$ and $\delta^{\mathrm{f}}_p$:
\begin{align*}
	\mathcal{P}\left( \{ \delta^{\mathrm{f}}_p \} , \{ \delta_p^{\mathrm{ic}} \} \right) &= \Pi\left( \{ \delta_p^{\mathrm{ic}} \} \right) \mathcal{P}\left( \{ \delta^{\mathrm{f}}_p \} | \{ \delta_p^{\mathrm{ic}} \} \right) \\
    &= \Pi\left( \{ \delta_p^{\mathrm{ic}} \} \right) \prod_p \delta^{\mathrm{D}} \left[ \delta^{\mathrm{f}}_p - \mathcal{M}_p \left( \{ \delta_p^{\mathrm{ic}} \} \right) \right] ,
    \numberthis
	\label{eq:joint_prior_final_density_altair}
\end{align*}
after plugging in the conditional posterior distribution from Eq.~\eqref{eq:conditional_posterior_final_density_altair}.

Assuming a normal distribution with zero mean and covariance $\mymat{S}$ corresponding to an underlying cosmological power spectrum for the initial conditions $\delta_p^{\mathrm{ic}}$, the joint prior distribution can be expressed as
\begin{align*}
	\mathcal{P}\left( \{ \delta^{\mathrm{f}}_p \} , \{ \delta_p^{\mathrm{ic}} \} |\, \mymat{S} \right) &= \Pi \left( \{ \delta_p^{\mathrm{ic}} \} |\, \mymat{S} \right) \prod_p \delta^{\mathrm{D}} \left[ \delta^{\mathrm{f}}_p - \mathcal{M}_p \left( \{ \delta_p^{\mathrm{ic}} \} \right) \right] \\
    &= \frac{\mathrm{e}^{- \frac{1}{2} \sum_{p,q} \delta_p^{\mathrm{ic}} \mymat{S}^{-1}\delta_q^{\mathrm{ic} }}} {\mathrm{det}(2 \pi\, \mymat{S})} \prod_p \delta^{\mathrm{D}} \left[ \delta^{\mathrm{f}}_p - \mathcal{M}_p \left( \{ \delta_p^{\mathrm{ic}} \} \right) \right] . \numberthis
	\label{eq:conditional_joint_prior_final_density_altair}
\end{align*}

Reformulating the statistical inference problem in terms of the initial conditions $\delta_p^{\mathrm{ic}}$ results in the following joint posterior distribution of $\delta_p^{\mathrm{ic}}$ and $\delta_p^{\mathrm{f}}$, conditional on the observed galaxy number counts $N^g_p$,
\begin{equation}
	\mathcal{P} \left( \{ \delta_p^{\mathrm{f}} \}, \{ \delta_p^{\mathrm{ic}} \} | \{ N^g_p \}, \mymat{S} \right) = \frac{\mathcal{L}\left( \{ N^g_p \} | \{ \delta_p^{\mathrm{f}} \}, \{ \delta_p^{\mathrm{ic}} \} \right) \mathcal{P}\left( \{ \delta_p^{\mathrm{f}} \}, \{ \delta_p^{\mathrm{ic}} \} | \, \mymat{S} \right)}{\Pi\left( \{ N^g_p \} | \, \mymat{S} \right)} ,
	\label{eq:derivation_density_posterior_altair}
\end{equation}
after making the dependence on the underlying power spectrum explicit. Assuming that the galaxy observations are conditionally independent of the initial conditions once the final density field is known, i.e. $\mathcal{L}( \{ N^g_p \} | \{ \delta_p^{\mathrm{f}} \}, \{ \delta_p^{\mathrm{ic}} \} ) = \mathcal{L}( \{ N^g_p \} | \{ \delta_p^{\mathrm{f}} \} )$ and using the joint prior distribution from Eq.~\eqref{eq:conditional_joint_prior_final_density_altair} leads to the LSS posterior distribution:
\begin{multline}
	\mathcal{P} \left( \{ \delta_p^{\mathrm{f}} \}, \{ \delta_p^{\mathrm{ic}} \} | \{ N^g_p \}, \mymat{S} \right) = \mathcal{L}\left( \{ N^g_p \} | \{ \delta_p^{\mathrm{f}} \} \right) \frac{\Pi \left( \{ \delta_p^{\mathrm{ic}} \} |\, \mymat{S} \right)}{\Pi\left( \{ N^g_p \} |\, \mymat{S} \right)} \\ \times \prod_p \delta^{\mathrm{D}} \left[ \delta^{\mathrm{f}}_p - \mathcal{M}_p \left( \{ \delta_p^{\mathrm{ic}} \} \right) \right] .
	\label{eq:lss_posterior_altair}
\end{multline}
By marginalizing over the final density field $\delta^{\mathrm{f}}_p$, we finally obtain our posterior distribution as follows:
\begin{equation}
	\mathcal{P} \left( \{ \delta_p^{\mathrm{ic}} \} | \{ N^g_p \}, \mymat{S} \right) = \mathcal{L}\left[ \{ N^g_p \} \big| \mathcal{M}_p \left(\{ \delta_p^{\mathrm{ic}} \}\right) \right] \frac{\Pi\left( \{ \delta_p^{\mathrm{ic}} \} |\, \mymat{S}\right) }{\Pi\left( \{ N^g_p \} | \, \mymat{S} \right)} , 
	\label{eq:final_density_posterior_altair}
\end{equation} 
thereby connecting present galaxy observations $N^g_p$ to their corresponding primordial density fields $\delta_p^{\mathrm{ic}}$ via a given forward model, $\mathcal{M}_p ( \{ \delta_p^{\mathrm{ic}} \})$. We must therefore sample from the highly non-linear and non-Gaussian posterior above to obtain realizations of the 3D initial density fields conditioned on the galaxy observations via the sophisticated HMC method described below in Appendix \ref{hmc}.

\subsection{Choice of density prior}
\label{choice_density_prior}

Standard Wiener filtering approaches employ isotropic Gaussian priors on the present-day density contrast which is justified for inference on the largest scales \citep[e.g.][]{zaroubi1999wiener, zaroubi2002unbiased, erdogdu20042df, erdogdu2006reconstructed, kitaura2008bayesian, kitaura2009cosmic, jasche2010bayesian, jasche2015matrix}, where the density field can be reasonably approximated as a Gaussian random field. Although the exact probability distribution for the density field in the mildly and strongly non-linear regimes is not known, the deviation from Gaussianity is well-established and we therefore require a non-Gaussian prior to effectively capture the details of the highly complex filamentary nature of the present day cosmic web.

Lognormal density priors were subsequently proposed \citep{coles1991lognormal} and this proved to be an adequate phenomenological choice \citep[e.g.][]{hubble1934distribution, peebles1980large, gaztanaga1993probing, kayo2001probability}, albeit with some limitations, for modelling the evolved matter field in the mildly non-linear regime \citep{kitaura2009cosmic, jasche2010fast,jasche2010bayesian_nonlinear}. The use of Edgeworth expansions to construct prior distributions involving third order moments has also been proposed in the literature \citep{kitaura2012nongaussian}.

In this work, we rely on Lagrangian Perturbation Theory (LPT) to model cosmic structure formation, which accounts for non-local effects of gravitational mass transport from initial to final positions. It has been extensively shown that LPT provides a sufficient description of the cosmic LSS on large scales, where it is capable of reproducing the exact one-, two- and three-point statistics of the evolved field, while still being a reasonable approximation to the higher order statistics \citep[e.g.][]{moutarde1991precollapse, buchert1994testing, bouchet1995perturbative, scoccimarro2000bispectrum, scoccimarro2002pthalos}. 

The essence of the above approach is that it provides a pathway to recover the high-order statistics of the matter distribution using only the 2-point statistics of the initial conditions, obviating the need for additional parameters to describe higher order statistics for the matter inference problem. Moreover, since our model encodes the dynamics, the reconstruction of the large-scale velocity field is automatically generated without incurring the expense of an increased parameter space.

\subsection{Modular statistical programming}
\label{block_sampling}

\begin{figure*}
	\centering
		{\includegraphics[scale=0.55]{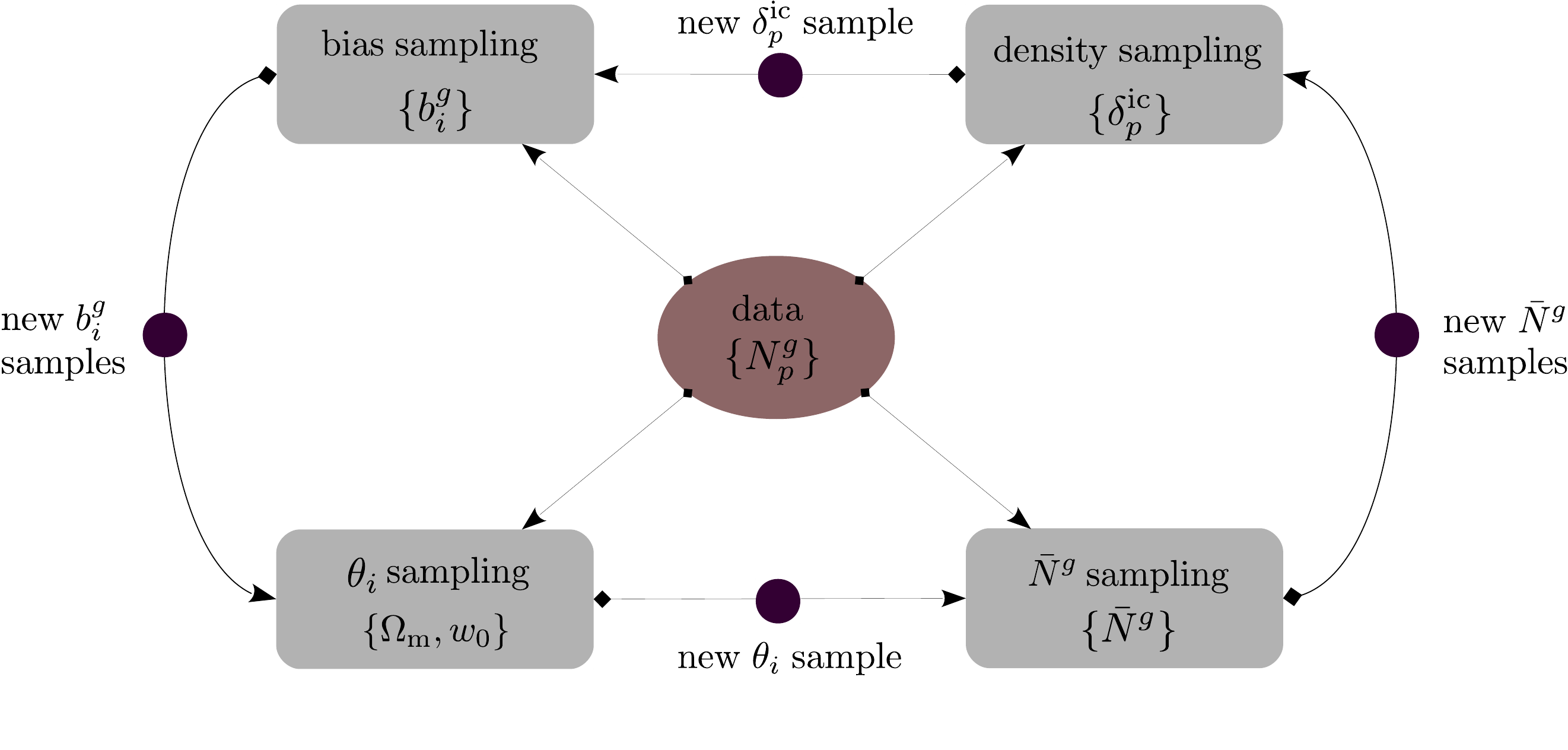}}
	\caption{Schematic representation of the multi-step iterative block sampling procedure implemented in \textsc{altair}. Initially, a realization of the 3D density contrast is obtained conditional on galaxy observations, followed by the other parameters being sampled conditional on the respective previous samples. Subsequent iterations of this procedure yield samples from the full joint posterior distribution within a modular statistical programming framework.}
	\label{fig:block_sampling_figure}
\end{figure*}

In this work, we exploit the modular statistical programming framework inherent in the \textsc{borg} algorithm to encode the AP test as an additional component to the original block sampling machinery. This approach allows us to model any Bayesian hierarchical problem to take into account any observational systematics via data models of higher complexity. Here, we account for the unknown parameters $\{b_i^g\}$ of the galaxy bias model described in Section \ref{poisson_likelihood} and unknown normalizations $\{\bar{N}^g\}$ for distinct galaxy samples, as illustrated in Fig. \ref{fig:block_sampling_figure}. These last normalizations, in practice, scale as the noise amplitudes as we are using a per-voxel Poissonian likelihood.

Conceptually, within such a framework, the overall aim is to characterize fully the augmented joint posterior $\mathcal{P} (\{\delta^{\mathrm{ic}}_p\}, \{\bar{N}^g\}, \{b_i^g\}, \{\theta_i\} |\{N^g_p\}, \mymat{S})$ of different tracer populations labelled by the index $g$. Since it is not computationally feasible and advisable to sample directly from the high dimensional joint posterior distribution, we make use of an important theorem on Metropolis-Hastings block sampling which allows us to break the high dimensional sampling problem into a series of lower dimensional ones \citep{hastings1970monte}. We therefore dissect the exploration of the full joint parameter space into a sequence of conditional sampling procedures. The block sampling approach consists of drawing samples from the following conditional probability distributions:
\begin{align*}
	&(1) \: \{\delta^{\mathrm{ic}}_p\}^{s+1} \curvearrowleft \mathcal{P} \left(\{\delta^{\mathrm{ic}}_p\} | \{\theta_i\}^s, \{N^g_p\}^s, \{\bar{N}^g\}^s, \{b_i^g\}^s, \mymat{S} \right) \\
	&(2) \: \{\bar{N}^g\}^{s+1} \curvearrowleft \mathcal{P} \left(\{\bar{N}^g\} | \{\theta_i\}^s, \{N^g_p\}^s, \{\delta^{\mathrm{ic}}_p\}^s, \{b_i^g\}^s, \mymat{S} \right) \\
	&(3) \: \{b_i^g\}^{s+1} \curvearrowleft \mathcal{P} \left(\{b_i^g\} | \{\theta_i\}^s, \{\bar{N}^g\}^s, \{N^g_p\}^s, \{\delta^{\mathrm{ic}}_p\}^s, \mymat{S} \right) \\
	&(4) \: \{\theta_i\}^{s+1} \curvearrowleft \mathcal{P} \left(\{\theta_i\} | \{\bar{N}^g\}^s, \{N^g_p\}^s, \{\delta^{\mathrm{ic}}_p\}^s, \{b_i^g\}^s, \mymat{S} \right). \numberthis \label{eq:block_sampling_altair}
\end{align*}
In the above expressions, $s$ denotes the sampling step and the symbol $\curvearrowleft$ indicates sampling from the distribution on the right hand side. A series of iterations of the individual sampling steps above will converge to the target distribution, i.e. the full joint posterior distribution \citep{hastings1970monte}. Hence, by simultaneously exploring the 3D initial density field $\{\delta^{\mathrm{ic}}_p\}$, the galaxy bias parameters $\{b_i^g\}$, the galaxy density contrast normalizations $\{\bar{N}^g\}$ and the cosmological parameters $\{\theta_i\}$ via an implementation of high dimensional MCMC methods in a multiple block sampling procedure, we can obtain samples from the desired joint posterior distribution.

\subsection{The bias posterior distribution}
\label{bias_posterior}

The formalism for the data model is presented in a generic context in Section \ref{poisson_likelihood}, such that it can be implemented for two or more different galaxy surveys, distinct subsamples of a given catalogue or even different cosmological probes of the LSS. The advantage of splitting a galaxy sample into various subsamples is that we can treat the respective systematic and statistical uncertainties of each subsample separately, thereby allowing us to account for the distinct clustering behaviour of galaxy populations in the LSS sample via their respective bias parameters.

These additional parameters can be trivially incorporated in the flexible block sampling approach adopted here, as described in Section \ref{block_sampling}. The mean numbers of galaxies, $\bar{N}^g$, for the various subsamples are essential for defining the density contrast of the galaxy distribution, with an erroneous value of $\bar{N}^g$ resulting in a non-zero value of the mean, yielding spurious large-scale power in the inferred density fields \citep{jasche2013methods}. Due to a lack of a priori knowledge, we must perform a joint inference of the set of four $\bar{N}^g$ and $b_i^g$ bias parameters, together with initial and final density fields, to take into account possible cross-correlations and interdependencies. Unlike traditional approaches \citep[e.g.][]{tegmark2004threedimensional}, here we infer the bias factors directly from the relation between the data and the density field.

The conditional posterior distribution for the bias parameters, given a realization of the density field and galaxy number counts for the respective subcatalogues, can be expressed as:
\begin{multline}
	\mathcal{P} \left( \{\bar{N}^g\}, \{b_i^g\} | \{ \theta_i \}, \{N^g_p\}, \{ \delta_p^{\mathrm{ic}} \} \right) \\ \propto \mathcal{L}\left[\{N^g_p\}\big| \mathcal{M}_p \left(\{ \delta_p^{\mathrm{ic}} \} \right), \{ \theta_i \}, \{\bar{N}^g\}, \{b_i^g\} \right] \times \Pi\left( \{\bar{N}^g\}, \{b_i^g\} \right),
	\label{eq:bias_posterior_altair}
\end{multline}
following an analogous reasoning to that described in Section \ref{theta_posterior}. Adopting a standard maximally agnostic philosophy, we set uniform prior distributions for the bias parameters:
\begin{equation}
	\Pi\left( \{\bar{N}^g\}, \{b_i^g\} \right) = \Theta\left( \{\bar{N}^g\} \right) \Theta\left( \{b_i^g\} \right) ,
	\label{eq:bias_prior_altair}
\end{equation}
where the Heaviside function $\Theta(x)$ ensures that the parameters are positive, as required by the bias model. The non-linear shape of the galaxy biasing function, as given in Eq.~\eqref{eq:poisson_intensity_biased_altair}, poses 
a particular challenge as no straightforward sampling procedure can be derived. We therefore sample from the above conditional bias posterior distribution via sequential slice sampling steps \citep[e.g.][]{neal2000slicesampling, neal2003slice} to ensure unit acceptance rates.

\section{Jacobian of comoving-redshift transformation}
\label{jacobian}

The derivation of the Jacobian, $\bm{\mathcal{J}}^z_r$, of the coordinate transformation between comoving $(\myvec{r})$ and redshift $(\myvec{z})$ transformation is described below.

Since we desire the redshift space representation of the density field, 
\begin{equation}
	\rho_z [\myvec{z}(\myvec{r})] = \rho_r(\myvec{r}) \left| \frac{\partial z}{\partial r} \right|^{-1} = \rho_r(\myvec{r}) \big| \bm{\mathcal{J}}^r_z \big|^{-1} = \rho_r(\myvec{r}) \big| \bm{\mathcal{J}}^z_r \big| ,
	\label{eq:rho_jacobian_altair}
\end{equation} 
or in terms of the density contrast,
\begin{equation}
	\delta_z [\myvec{z}(\myvec{r})] = [1 + \delta_r(\myvec{r})] \big| \bm{\mathcal{J}}^z_r \big| - 1 ,
	\label{eq:density_contrast_jacobian_altair}
\end{equation} 
where $\bm{\mathcal{J}}^r_z$ is the corresponding Jacobian for the converse (redshift to comoving) transformation.  

The set of three indices $\{ i,j,k \}$ below are spatial coordinates $\{ 1, 2, 3 \}$. With the functional dependence of $\myvec{z} = f(\myvec{r})$, we have 
\begin{equation}
	z_i = \frac{r_i}{r} f(\myvec{r}) = \bar{z}\left(r\right) \frac{r_i}{r},
	\label{eq:functional_dependence_z2r_altair}
\end{equation} 
where we defined $r = \sqrt{\sum_j r_j^2}$, and $\bar{z}$ is the cosmological redshift. Taking the derivative yields
\begin{align*}
	\mathrm{d}z_i &= \mathrm{d}\bar{z}\left(r\right) \frac{r_i}{r} + \bar{z}\left(r\right) \mathrm{d}\left( \frac{r_i}{r} \right) \\
	&= \sum_j \frac{\partial \bar{z}}{\partial r} \frac{\partial r}{\partial r_j} \mathrm{d}r_j \frac{r_i}{r} + \bar{z} \frac{\mathrm{d}r_i}{r} + \bar{z} r_i \mathrm{d} \left( \frac{1}{r} \right) ,  \numberthis
	\label{eq:derivative1_jacobian_altair}
\end{align*}
after applying the chain rule. Differentiating with respect to the comoving coordinate,
and using $\mathrm{d} \left( 1 / r \right) = -\sum_j (1/r^3) r_j \mathrm{d}r_j$, 
\begin{align*}
	\frac{\partial z_i}{\partial r_k} &= \frac{\partial \bar{z}}{\partial r} \frac{\partial r}{\partial r_k} \frac{r_i}{r} + \frac{\bar{z}}{r} \delta^{\mathrm{k}}_{ik} - \frac{z r_i r_k}{r^3} \\
	&= \left( \frac{\partial \bar{z}}{\partial r} - \frac{\bar{z}}{r} \right) \frac{r_i r_k}{r^2} + \frac{\bar{z}}{r} \delta^{\mathrm{k}}_{ik} ,  \numberthis
	\label{eq:derivative2_jacobian_altair}
\end{align*}
where we made use of $r {\mathrm{d} r} = \sum_k r_k {\mathrm{d} r_k}$ which follows from the definition $r = \sqrt{\sum_k r_k^2}$. To compute $\partial \bar{z} / \partial r$, we start from the definition of comoving distance, 
\begin{equation}
	r = \frac{c}{H_0} \int_0^{\bar{z}} \frac{\mathrm{d} z}{\mathpzc{E}(z)} ,
	\label{eq:comoving_distance_definition_altair}
\end{equation} 
with the conventional definition for $\mathpzc{E}(z)$,
\begin{equation}
	\mathpzc{E}(z) = \sqrt{\Omega_{\mathrm{de}} (1 + z)^{3(1 + w)} + \Omega_{\mathrm{k}}(1 + z)^2 + \Omega_{\mathrm{m}}(1 + z)^3 + \Omega_{\mathrm{r}}(1 + z)^4} ,
	\label{eq:E_z_comoving_distance_definition_altair}
\end{equation} 
where $c$ is the speed of light and $H_0$ is the Hubble constant today, such that 
\begin{equation}
	\frac{\partial \bar{z}}{\partial r} = \left( \frac{H_0}{c} \right) \mathpzc{E}(\bar{z}) .
	\label{eq:derivative_comoving_distance_definition_altair}
\end{equation} 
Finally, the Jacobian matrix can be expressed as follows:
\begin{equation}
	{\mathcal{J}}^r_{z\{ ik \}} = \frac{\partial z_i}{\partial r_k} =  \left( \frac{H_0}{c} \mathpzc{E}(\bar{z}) - \frac{\bar{z}}{r} \right) \frac{r_i r_k}{r^2} + \frac{\bar{z}}{r} \delta^{\mathrm{k}}_{ik} ,
	\label{eq:final_expression_jacobian_altair}
\end{equation}
and we obtain the desired $\big| {\mathcal{J}}^z_{r\{ ik \}} \big|$ by taking the reciprocal of $\big| {\mathcal{J}}^r_{z\{ ik \}} \big|$. 

\section{Hamiltonian sampling}
\label{hmc}

The exploration of the LPT-Poissonian posterior, described by Eq.~\eqref{eq:joint_posterior_altair}, requires highly non-linear reconstruction methods and is therefore numerically intensive. From the Bayesian viewpoint considered in this work, a single estimate of the density field is not of particular interest, with the desired scientific output being a sampled representation of the multidimensional LSS posterior. Extracting any relevant statistical summary, such as mean, mode or variance, given this posterior representation, is then straightforward. Furthermore, the propagation of uncertainties to the final inferred quantities is smooth and coherent.  

In the absence of a known procedure to directly sample from a LPT-Poissonian distribution, a Metropolis-Hastings sampling scheme with an accept-reject step must be implemented. But such a mechanism suffers from the well-known drawback of possible high rejection rate, where a low acceptance rate of proposed samples will render the method numerically inefficient. This is especially significant since the inference of 3D density fields typically involves extremely high number of inference parameters. For instance, in this work, there are over $2 \times 10^6$ free parameters, which correspond to primordial density fluctuation amplitudes $\delta^{\mathrm{ic}}_p$ at respective grid nodes. Exploring this high dimensional parameter space via a random walk, as in conventional Markov Chain Monte Carlo (MCMC) methods, consequently results in a high rejection rate. This is an understatement, as in practice, the acceptance rate is virtually zero, with one chance over $10^6$ of going in the right direction, without taking into account the step size.

To limit the number of samples and alleviate the numerical scaling of the method, we implement a Hamiltonian Monte Carlo (HMC) scheme. This method guarantees an acceptance rate of unity in the absence of numerical errors. The numerical efficiency of the Hamiltonian sampling lies in deterministically proposing new samples to the Metropolis-Hastings algorithm, based on techniques to follow dynamical particle motion in potentials. The HMC algorithm therefore maintains a reasonable sampling efficiency in high dimensional spaces by suppressing the dominant random walk behaviour.

The HMC method has been successfully implemented for the inference of density fields in the non-linear regime, and has been found to be very efficient \citep[e.g.][]{jasche2010fast, jasche2010bayesian_nonlinear, kitaura2012multiscale, jasche2013bayesian, jasche2013methods}. Moreover, this approach is less prone to spurious effects induced by numerical inaccuracies due to the final accept/reject step of the Metropolis-Hastings sampler ensuring correctness of the target density. Numerical inaccuracies would only be detrimental to the efficiency, without comprimising the correctness. 

In the following section, we briefly review the underlying framework and rationale of the HMC method. An excellent in-depth review of this approach is provided in \cite{duane1987hybrid} and \cite{neal1993probabilistic}, with its application to LSS inference illustrated in \cite{jasche2010fast}, \cite{jasche2010bayesian_nonlinear}, \cite{jasche2013bayesian, jasche2013methods} and \cite{jasche2018physical}.

\subsection{The HMC algorithm}
\label{hmc_formalism}

The essence of the HMC method is as follows: If we want to generate random variates according to a desired probability distribution $\mathcal{P}(\{ x_i \})$, with $\{ x_i \}$ being a set of $N$ elements $x_i$, then we may interpret the negative logarithm of the distribution as a potential $\psi(x)$,
\begin{equation}
	\psi(x) = - \ln (\mathcal{P}(x)).
	\label{eq:HMC_psi_potential_altair}
\end{equation}
We can subsequently formulate a corresponding Hamiltonian that describes the dynamics in the multidimensional phase space by adding a kinetic term to the above potential. To this end, we introduce a ``momentum'' variable $p_i$ and a ``mass matrix'' $\mymat{M}$ as nuisance parameters, as follows:
\begin{equation}
	H = \frac{1}{2} \sum_{i,j} p_i M_{ij}^{-1} p_j + \psi(x).
	\label{eq:HMC_hamiltonian_altair}
\end{equation}
This Hamiltonian, as in classical mechanics, describes the dynamics in a high dimensional parameter space. We then obtain the new target probability distribution by exponentiating the above Hamiltonian:
\begin{equation}
	e^{-H} = \mathcal{P}(\{ x_i \}) \exp{\Bigg(\frac{1}{2} \sum_{i,j} p_i M_{ij}^{-1} p_j  \Bigg)}.
	\label{eq:HMC_target_distribution_altair}
\end{equation}
The form of the Hamiltonian ensures that the new joint distribution can be separated into a Gaussian distribution for the momenta $\{ p_i \}$ and the target distribution $\mathcal{P}(\{ x_i \})$. This implies that the two sets of variables $\{ x_i \}$ and $\{ p_i \}$ are independent, and hence, we can obtain samples from the target distribution $\mathcal{P}(\{ x_i \})$ simply by marginalizing over the auxiliary momenta. 

The next step is to generate a random variate from the joint distribution, which is proportional to $\exp{(-H)}$. We must first draw a set of momenta from the distribution defined by the kinetic energy term, which is an $N$-dimensional Gaussian with covariance $\mymat{M}$. To obtain a new sample from the joint distribution, we allow the system to evolve deterministically, from a given initial point in the high dimensional parameter space $(\{ x_i \}, \{ p_i \})$, according to Hamilton's equations:
\begin{align}
	\frac{\mathrm{d} x_i}{\mathrm{d} t} &= \frac{\partial H}{\partial p_i} 
	\label{eq:HMC_hamilton_1st_equation_altair} \\
	\frac{\mathrm{d} p_i}{\mathrm{d} t} &= \frac{\partial H}{\partial x_i} = - \frac{\partial \psi(x)}{\partial x_i}. 
	\label{eq:HMC_hamilton_2nd_equation_altair}
\end{align}
Integrating the above equations of motion for the time variable $t$ up to a fixed pseudo-time $\tau$ results in the new position $(\{ x'_i \}, \{ p'_i \})$ in phase space. This new point in phase space is accepted according to the standard Metropolis-Hastings acceptance rule:
\begin{equation}
	\mathcal{P}_{\mathpzc{A}} = \mathrm{min}\left\lbrace 1, \exp{\left[ -\left(H\left(\{ x'_i \}, \{ p'_i \}\right) - H\left(\{ x_i \}, \{ p_i \}\right) \right) \right]}  \right\rbrace .
	\label{eq:HMC_metropolis_hastings_acceptance_altair}
\end{equation}
Since the equations of motion provide a solution to a Hamiltonian system, the energy or the Hamiltonian described by Eq.~\eqref{eq:HMC_hamiltonian_altair} must be conserved. A direct consequence of this conservation of the Hamiltonian is that the HMC approach yields an acceptance rate of unity. In practice, however, the acceptance rate may be lower due to numerical inaccuracies in the integration scheme. After accepting a new sample, the mechanism proceeds by discarding the momentum counterpart and restarts the sampling procedure by randomly drawing a new set of momenta. The individual momenta $\{ p_i \}$ do not have to be stored in memory which is numerically convenient, and marginalization simply implies discarding them. 

The HMC sampling scheme can therefore be summarized in two steps. The first step is essentially a Gibbs sampling step which results in a new set of Gaussian distributed momenta. The second step involves computing the deterministic dynamical trajectory on the posterior surface. We obtain samples of the desired target distribution by marginalizing over the nuisance parameters, i.e. by discarding the auxiliary momenta $\{ p_i \}$.

\section{Equations of motion for LSS inference}
\label{equations_of_motion}

As described in Appendix \ref{hmc}, the Hamiltonian Monte Carlo (HMC) approach allows the exploration of the non-linear LSS posterior via the study of the Hamiltonian dynamics in the high dimensional parameter space. To ensure a high acceptance rate, the HMC exploits the gradient of the logarithmic posterior distribution to optimally explore this parameter space, such that the algorithm also requires the derivatives of the posterior distribution, as outlined below. 

From the LSS posterior illustrated in Eq.~\eqref{eq:joint_posterior_altair}, we can derive the corresponding forces required to evaluate the Hamiltonian trajectories. The Hamiltonian potential $\psi(\delta^{\mathrm{ic}})$ can be written as 
\begin{align*}
	\psi(\delta^{\mathrm{ic}}) &= - \ln \left[ \mathcal{P} \left(\{\delta^{\mathrm{ic}}_p\}, \{\bar{N}^g\}, \{b^g\}, \{\theta_i\} |\{N^g_p\}, \mymat{S} \right) \right] \\ &= \psi_{\mathrm{likelihood}}(\delta^{\mathrm{ic}}) + \psi_{\mathrm{prior}}(\delta^{\mathrm{ic}}) , \numberthis
	\label{eq:psi_sum_altair}
\end{align*} 
where the potential $\psi_{\mathrm{prior}}(\delta^{\mathrm{ic}})$ corresponds to
\begin{equation}
	\psi_{\mathrm{prior}}(\delta^{\mathrm{ic}}) = \frac{1}{2} \delta^{\mathrm{ic}, \dagger} \mymat{S}^{-1} \delta^{\mathrm{ic}} = \frac{1}{2} \sum_{ij} \delta^{\mathrm{ic},*}_i S_{ij}^{-1} \delta^{\mathrm{ic}}_j ,
	\label{eq:psi_prior_altair}
\end{equation}
while $\psi_{\mathrm{likelihood}}(\delta^{\mathrm{ic}})$ is given by
\begin{multline}
	\psi_{\mathrm{likelihood}}(\delta^{\mathrm{ic}}) = \sum_p \Bigg\lbrace  \lambda^g_p \left( \{ \delta_p^{\mathrm{f}} \}, \{ \theta_i \}, \{ \bar{N}^g \}, \{ b^g_i \} \right) \\ - N^g_p \ln \left[ \lambda^g_p \left( \{ \delta_p^{\mathrm{f}} \}, \{ \theta_i \}, \{ \bar{N}^g \}, \{ b^g_i \} \right) \right] \Bigg\rbrace . \numberthis 
	\label{eq:psi_likelihood_altair}
\end{multline}
Using the definition of the Hamiltonian potential $\psi(\delta^{\mathrm{ic}})$ from Eq.~\eqref{eq:psi_sum_altair}, we obtain the corresponding equations describing the Hamiltonian forces by differentiating with respect to $\delta^{\mathrm{ic}}$:
\begin{equation}
	\frac{\partial \psi(\delta^{\mathrm{ic}})}{\partial \delta^{\mathrm{ic}}_i} = \frac{\partial \psi_{\mathrm{prior}}(\delta^{\mathrm{ic}})}{\partial \delta^{\mathrm{ic}}_i} + \frac{\partial \psi_{\mathrm{likelihood}}(\delta^{\mathrm{ic}})}{\partial \delta^{\mathrm{ic}}_i}.
	\label{eq:psi_derivative_def_altair}
\end{equation} 
The prior term is trivially obtained as
\begin{equation}
	\frac{\partial \psi_{\mathrm{prior}}(\delta^{\mathrm{ic}})}{\partial \delta^{\mathrm{ic}}_i} = \sum_j \delta^{\mathrm{ic,*}}_j S^{-1}_{ij} .
	\label{eq:psi_derivative_prior_altair}
\end{equation} 
The corresponding likelihood term, however, cannot be derived in straightforward fashion. We compute this adjoint gradient as a sequential application of linear operations, as follows:
\begin{equation}
	\frac{\partial \psi_{\mathrm{likelihood}}(\delta^{\mathrm{ic}})}{\partial \delta^{\mathrm{ic}}_s} = \sum\limits_{q} \frac{\partial \rho^{(r)}_q}{\partial \delta^{\mathrm{ic}}_s} \sum\limits_{\tilde{p}} \frac{\partial \rho^{(z)}_{\tilde{p}}}{\partial \rho^{(r)}_q} \sum\limits_{p,g}\frac{\partial \lambda^g_p}{\partial \rho^{(z)}_{\tilde{p}}} \frac{\partial \psi_{\mathrm{likelihood}}}{\partial \lambda^g_p} ,
	\label{eq:psi_derivative_likelihood_altair}
\end{equation} 
where we explicitly expressed the coordinate transformation from comoving $(\myvec{r})$ to redshift $(\myvec{z})$ space, and $\rho_p = 1 + \delta^{\mathrm{f}}_p$ is the density field. Eq.~\eqref{eq:psi_derivative_likelihood_altair} constitutes a sequence of matrix vector applications, as follows:
\begin{align}
	\Gtilde_{qs} &\equiv \frac{\partial \rho^{(r)}_q} {\partial \delta^{\mathrm{ic}}_s} \\
	\mathcal{Q}_{\tilde{p}q} &\equiv \frac{\partial \rho^{(z)}_{\tilde{p}}}{\partial \rho^{(r)}_q} = \sum_{i} \mathcal{E}^{-1}_{i\tilde{p}} \sum_{\tilde{p}} \mathcal{J}_{\tilde{p}}  x^{\alpha (i)}_{\tilde{p}} y^{\beta (i)}_{\tilde{p}} z^{\gamma (i)}_{\tilde{p}} \\
	\mathcal{K}_{p \tilde{p}} &\equiv \sum_g \frac{\partial \lambda^g_p}{\partial \rho^{(z)}_{\tilde{p}}} = \sum_g \delta^{\mathrm{k}}_{p,\tilde{p}} \frac{\lambda^g_p}{ \rho^{(r)}_{\tilde{p}} } \left[ \rho^g \epsilon^g \rho^{-\epsilon_g,(r)}_{\tilde{p}} + \beta \right] \\
	\mathpzc{v}_p &\equiv \frac{\partial \psi_{\mathrm{likelihood}}}{\partial \lambda^g_p} =  1 -  \frac{N^g_p}{\lambda^g_p \left( \{ \delta_p^{\mathrm{f}} \}, \{ \theta_i \}, \{ \bar{N}^g \}, \{ b^g_i \} \right)} ,
	\label{eq:adjoint_gradient_term_altair}
\end{align}
where $\mathcal{J}$ is the Jacobian of the comoving-redshift transformation, $\Gtilde$ indicates the derivative of LPT (cf. Appendix D in \cite{jasche2013bayesian}) and $\delta^{\mathrm{k}}_{p,\tilde{p}}$ is the Kronecker delta. The derivation of the adjoint gradient for the triquintic interpolation is illustrated in Appendix \ref{adjoint_interpolation}.

Using the above results, the two equations of motion \eqref{eq:HMC_hamilton_1st_equation_altair} and \eqref{eq:HMC_hamilton_2nd_equation_altair} for the Hamiltonian system can be expressed as follows:
\begin{equation}
	\frac{\mathrm{d} \hat{\delta}^{\mathrm{ic}}_i}{\mathrm{d} t} = \sum_j M_{ij}^{-1} p_j ,
	\label{eq:altair_1st_equation_altair}
\end{equation}
and	
\begin{equation}
	\frac{\mathrm{d} \hat{p}_i}{\mathrm{d} t} = - \sum_j \delta^{\mathrm{ic},*}_j S^{-1}_{ij} - \sum_{q,\tilde{p},p} \Gtilde_{qi} \mathcal{Q}_{\tilde{p}q} \, \mathcal{K}_{p\tilde{p}} \, \mathpzc{v}_p ,
	\label{eq:altair_2nd_equation_altair}
\end{equation}
where the hats denote Fourier space representation. We can now obtain new samples from the LSS posterior by following the dynamical evolution of the Hamiltonian system, governed by Eqs.~\eqref{eq:altair_1st_equation_altair} and \eqref{eq:altair_2nd_equation_altair}, in phase space.

\section{Numerical implementation}
\label{numerical_implementation}

Our numerical implementation of the augmented version of \textsc{borg}, that incorporates cosmological parameter inference, while assuming a more realistic non-linear bias model, is designated as \textsc{altair}. It employs the FFTW3 library for fast Fourier transforms \citep{frigo2005design}, whose feature of parallel transforms allows for straightforward parallelization of our code. For random number generation, we make use of the GNU scientific library (gsl) \citep{galassi2009gnu}, in particular, the Mersenne Twister MT$19937$, with $32$-bit word length, from the \texttt{gsl\_rng\_mt19937} routine. The use of the Mersenne Twister algorithm as a pseudo-random number generator for Monte Carlo simulations has been proven to be efficient and dependable \citep{matsumoto1998mersenne}.

\subsection{The leapfrog scheme}
\label{leapfrog_scheme}

In order to obtain samples from the LSS posterior, we must integrate the equations of motion \eqref{eq:altair_1st_equation_altair} and \eqref{eq:altair_2nd_equation_altair} numerically over a pseudo-time interval $\tau$ by means of a leapfrog scheme. Ideally, we want to solve the equations exactly for an optimal acceptance rate. As such, the choice of this integrator is motivated by several reasons. Primarily, the leapfrog scheme is convenient as it is a symplectic integrator and is therefore exactly reversible, thereby maintaining detailed balance \citep{duane1987hybrid} of the Markov chain. The high accuracy of such an integration scheme is conducive to the conservation of the total Hamiltonian along a given trajectory, within the limit of numerical errors, resulting in high acceptance rates. It is also numerically stable and allows for simple propagation of errors. To avoid resonant trajectories, $\tau$ is drawn from a uniform distribution. The numerical implementation of the ``kick-drift-kick'' leapfrog scheme here follows closely the descriptions provided in \citet{jasche2010fast} and \citet{jasche2013bayesian}, which we refer the reader to for more elaborate and complementary explanations.

\subsection{Hamiltonian mass}
\label{hamiltonian_mass}

The ``mass matrix'' $\mymat{M}$ contains a large number of parameters that must be tuned to optimize the performance of the HMC sampler. Conceptually, this Hamiltonian mass characterizes the inertia of individual parameters as they move through the parameter space. As a result, the choice of $\mymat{M}$ is a compromise to limit slow exploration efficiency due to extremely large masses and avoid large numerical errors of the integration scheme due to extremely light masses, and is therefore a trade-off between efficiency and acceptance rate. 

A general prescription for the Hamiltonian masses, when the density fluctuations $\delta^{\mathrm{f}}_p$ can be characterized as a Gaussian random field, is to set them inversely proportional to the variance of that specific $\delta^{\mathrm{f}}_p$ \citep{taylor2008fast}. However, this prescription has been found to be effective even for the case of non-Gaussian distributions, such as the LPT-Poissonian posterior encountered in this work \citep{jasche2013bayesian, jasche2015past, lavaux2016unmasking}. Since the leapfrog scheme requires the inversion of $\mymat{M}$, a diagonal representation of $\mymat{M}$ is numerically convenient, given the extremely high dimensionality of the problem. We therefore choose $\mymat{M}$ to be inversely proportional to the cosmological power spectrum, $\mymat{M} = \mymat{S}^{-1}$, i.e. diagonal in its Fourier basis. This choice, moreover, improves the efficiency of the HMC sampler and results in faster convergence since it accounts for the correlation structure of the underlying density field \citep{jasche2013bayesian}.

\section{Rotation of cosmological parameter space}
\label{rotation_cosmo}

\begin{figure*}
	\centering
		{\includegraphics[width=0.9\hsize,clip=true]{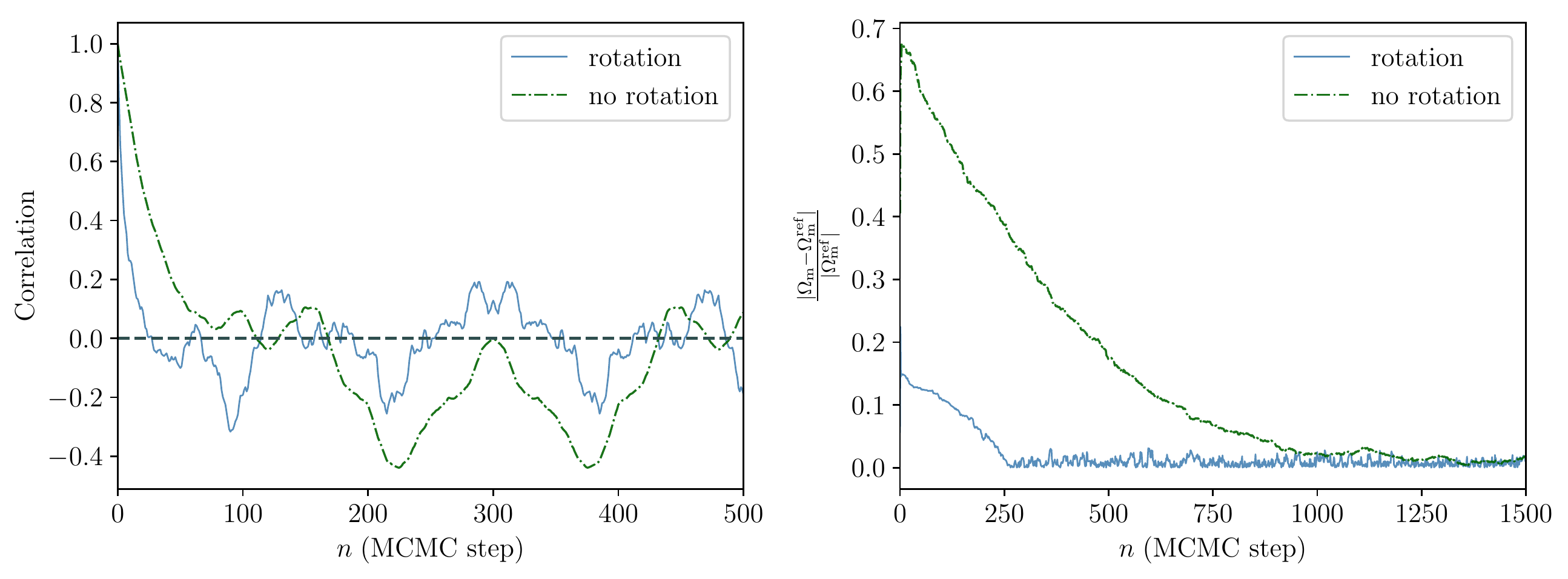}}
	\caption{{\it Left panel:} The correlation length of the $\Omega_{\mathrm{m}}$ Markov chain. The chain has a correlation length of the order of 25 samples. The dashed-dotted lines depict the longer correlation length without the rotation of the cosmological parameter space, by nearly a factor of five. {\it Right panel:} The burn-in phase for both scenarios, illustrated via the deviation from the reference cosmology. The dashed-dotted lines depict the MCMC chains obtained without rotating the parameter space, where the burn-in phase consequently requires around five times more iterations. The longer correlation length and burn-in phase highlight the loss in efficiency compared to the more sophisticated sampler.}
	\label{fig:correlation_length_cosmo_MCMC_chains_altair}
\end{figure*}

To increase the efficiency of the cosmological parameter sampler, we rotate the $( \Omega_{\mathrm{m}} , w_0 )$ parameter space, using orthogonal basis transformations, before slice sampling. The corresponding covariance matrix can be orthonormally decomposed as $\mymat{C} = \mymat{Q}^{\dagger} \bm{\Delta} \mymat{Q}$, where $\mymat{Q}^{\dagger} \mymat{Q} = \mathbb{1}$.

We perform a rotation of the $\bm{\theta}$ vector of cosmological parameters, about the mean $\bar{\bm{\theta}}$, using the transformation:
\begin{equation}
	\tilde{\bm{\theta}} = \bm{\Delta}^{-\frac{1}{2}} \mymat{Q} ( \bm{\theta} - \bar{\bm{\theta}} ) ,
	\label{eq:cosmo_rotation_appendix_altair}
\end{equation}
where the tilde denotes the transformed variable. A key point is that we must rotate back to the original frame when computing the logarithm of the posterior via
\begin{equation}
	\bm{\theta} = \mymat{Q}^{\dagger} \bm{\Delta}^{\frac{1}{2}}\tilde{\bm{\theta}} + \bar{\bm{\theta}} .
	\label{eq:cosmo_inverse_rotation_appendix_altair}
\end{equation}
This is implemented at each step of the Markov chain, such that the covariance used to compute the orthogonal basis operators must be learnt progressively as follows:
\begin{equation}
	\mymat{C}_{n + 1} = \frac{n}{n + 1} \mymat{C}_n + \frac{( \bm{\theta}_n - \bar{\bm{\theta}} )^{\dagger} ( \bm{\theta}_n - \bar{\bm{\theta}} )}{n}  ,
	\label{eq:adaptive_cosmo_covariance_appendix_altair}
\end{equation}
at the given $n$th step. The above transformation helps to decorrelate the variables involved when sampling, thereby reducing the burn-in period significantly. This procedure is especially effective with highly correlated parameters, as is the case for $( \Omega_{\mathrm{m}} , w_0 )$ (cf. Fig. \ref{fig:marginal_posteriors_cosmo_params_altair}).

We investigate the statistical efficiency of our cosmological parameter sampler by computing the correlation length of the Markov chain. Subsequent samples in the chain are generally correlated and hence do not qualify as independent samples of the target posterior distribution. The correlation length characterizes the statistical efficiency of generating independent samples of a given parameter $\theta$, as follows:
\begin{equation}
	C(\theta)_n = \frac{1}{N - n} \sum_{i=0}^{N - n} \left( \frac{\theta^i - \langle \theta \rangle}{\sqrt{\mathrm{Var}(\theta)}} \frac{\theta^{i + n} - \langle \theta \rangle}{\sqrt{\mathrm{Var}(\theta)}} \right) ,
	\label{eq:correlation_length_altair}
\end{equation}
where $n$ is the number of transition steps in the chain, $\langle \theta \rangle$ and $\mathrm{Var}(\theta)$ correspond to the mean and variance, respectively, i.e. $\langle \theta \rangle = 1/N \sum_i \theta^i$ and $\mathrm{Var}(\theta) = 1/N \sum_i ( \theta^i - \langle \theta \rangle )^2$, while $N$ is the total number of realizations in the chain. This correlation length therefore indicates the number of independent realizations that can be drawn from the MCMC chain with a given number of transition steps. For an unbiased comparison, we computed the correlation length using the same chain length for both scenarios, while ignoring their respective burn-in phases. From the left panel of Fig. \ref{fig:correlation_length_cosmo_MCMC_chains_altair}, we deduce that the correlation length for the $\Omega_{\mathrm{m}}$ chain is of the order of 25 samples. Without the rotation of $(\Omega_{\mathrm{m}} , w_0 )$ parameter space, the correlation length is longer by nearly a factor of five, further demonstrating the gain in efficiency obtained with the more sophisticated sampler.

\section{Adjoint gradient for generic 3D interpolation}
\label{adjoint_interpolation}

The 3D interpolation can be expressed generally, for any arbitrary $n$th order of interpolation, as:
\begin{equation}
	P_{\alpha \beta \gamma} = \sum_{i=0}^{n} \sum_{j=0}^{n} \sum_{k=0}^{n} a_{ijk} x_{\alpha}^i y_{\beta}^j z_{\gamma}^k = \sum_{i,j,k}^{n} a_{ijk} {\mathcal{E}_{\alpha \beta \gamma}}^{ijk},
	\label{eq:generic_3D_interpolation_appendix_altair}
\end{equation}
where $\mymat{P}$ is the interpolated surface and $\bm{\mathcal{E}}$ is a $(n+1)^3 \times (n+1)^3$ matrix required for computing the vector of interpolation coefficients $\myvec{a}$. In this work, we implement a triquintic interpolation scheme, which corresponds to $n = 5$, with the indices $\{ i,j,k \}, \{ \alpha, \beta, \gamma \} = \{ 0,1,2,3,4,5 \}$ constituting the particular indexing scheme employed to denote the voxels involved. $\bm{\mathcal{E}}$ is therefore a $216 \times 216$ matrix for the triquintic scheme.

The above system of equations can be reformulated as a matrix for the linear system described by $\mymat{P} = \bm{\mathcal{E}} \myvec{a}$, such that the vector of interpolation coefficients can be computed in straightforward fashion through matrix inversion, $\myvec{a} = \bm{\mathcal{E}}^{-1} \mymat{P}$. The advantage of this approach is that $\bm{\mathcal{E}}^{-1}$ can computed only once and stored, and then used for interpolation at any location inside the cube \citep[e.g.][]{lekien2005tricubic}.

The derivative of the generic 3D interpolation is required as a component of the adjoint gradient at the core of the HMC sampler described in Appendix~\ref{hmc}. The 3D interpolated density field, in redshift space (cf. Eq.~\eqref{eq:rho_jacobian_altair}), can be written explicitly as:
\begin{equation}
	\tilde{\rho}_k = \mathcal{J}_k \left( \sum_{i,m} \mathcal{E}_{im}^{-1} \rho_m x^{\alpha (i)}_k y^{\beta (i)}_k z^{\gamma (i)}_k \right) , 
	\label{eq:interpolated_z_field_appendix_altair}
\end{equation}
where $\bm{\mathcal{J}}$ is the Jacobian of the comoving-redshift transformation from Appendix \ref{jacobian}, evaluated at the location of the mesh element $k$. Here, we employed a compact notation, where $\{ x, y, z \}$ correspond to the fractional steps with respect to a reference node in the grid, i.e. interpolation weights, and $\{ \alpha, \beta, \gamma \}$ label the powers of the elements of $\bm{\mathcal{E}}$ via:
\begin{equation}
	i = \alpha (n + 1)^2 + \beta (n + 1) + \gamma ,
\end{equation}
with the index $m \in \{ i,j,k \}$ from Eq.~\eqref{eq:generic_3D_interpolation_appendix_altair}. Using the chain rule, the resulting gradient $\myvec{w}$, after application of the incoming gradient $\myvec{v}$ from the previous components in the forward model, is obtained as follows:
\begin{align*}
	w_a &= \sum_k v_k \frac{\partial \tilde{\rho}_k}{\partial \rho_a} \\ 
    &= \sum_k \mathcal{J}_k v_k \sum_{i,m} \mathcal{E}^{-1}_{im} \delta_{ma}^{\mathrm{k}} x^{\alpha (i)}_k y^{\beta (i)}_k z^{\gamma (i)}_k \\ 
    &= \sum_i \mathcal{E}^{-1}_{ia} \sum_k \mathcal{J}_k v_k x^{\alpha (i)}_k y^{\beta (i)}_k z^{\gamma (i)}_k \\
    &= \sum_{i,k} \mathcal{E}^{-1}_{ia} \xi_{ik} , \numberthis
	\label{eq:adjoint_gradient_interp3d_altair}
\end{align*}
where $\xi_{ik} \equiv \mathcal{J}_k v_k x^{\alpha (i)}_k y^{\beta (i)}_k z^{\gamma (i)}_k$ represents the contribution to the adjoint gradient from a specific voxel labelled by the indexing scheme adopted. This is convenient as $\bm{\mathcal{E}}^{-1}$ is already available from the execution of the 3D interpolation in the forward model.

\end{document}